\documentclass{aa}
\usepackage{graphicx}
\usepackage{natbib}
\usepackage[varg]{txfonts}
\usepackage{hyperref} 
\hypersetup{colorlinks=true,breaklinks=true,linkcolor=blue,citecolor=blue,urlcolor=blue}
\begin{document}
\title{On the (in)stability of sunspots\thanks{This paper is mainly based on Part\,I of the Ph.D. thesis ``On the decay of sunspots'', \url{https://freidok.uni-freiburg.de/data/165760}.}
}
\titlerunning{On the (in)stability of sunspots}
\authorrunning{Strecker et al.}
\author{H. Strecker\inst{1} \and W. Schmidt\inst{1} \and R. Schlichenmaier\inst{1} \and M. Rempel\inst{2}} 
\institute{%
 Leibniz-Institut f{\"u}r Sonnenphysik,
 Sch{\"o}neckstr.\ 6,
 79104 Freiburg,
 Germany,
 \email{strecker@leibniz-kis.de} 
\and 
 High Altitude Observatory, NCAR,
 P.O. Box 3000,
 Boulder, 
 CO 80307,
 USA
}
\date{Received: 22 December 2020 / Accepted: 04 March 2021}
\abstract
{The stability of sunspots is one of the long-standing unsolved puzzles in the field of solar magnetism and the solar cycle. The thermal and magnetic structure of the sunspot beneath the solar surface is not accessible through observations, thus processes in these regions that contribute to the decay of sunspots can only be studied through theoretical and numerical studies.} 
{We study the effects that destabilise and stabilise the flux tube of a simulated sunspot in the upper convection zone. The depth-varying effects of fluting instability, buoyancy forces, and timescales on the flux tube are analysed.}
{We analysed a numerical simulation of a sunspot calculated with the MURaM code. The simulation domain has a lateral extension of more than 98\,Mm\,$\times$\,98\,Mm and extends almost 18\,Mm below the solar surface. The analysed data set of 30\,hours shows a stable sunspot at the solar surface. We studied the evolution of the flux tube at defined horizontal layers (1) by means of the relative change in perimeter and area, that is, its compactness; and (2) with a linear stability analysis.}
{The simulation shows a corrugation along the perimeter of the flux tube (sunspot) that proceeds fastest at a depth of about 8\,Mm below the solar surface. Towards the surface and towards deeper layers, the decrease in compactness is damped. From the stability analysis, we find that above a depth of 2\,Mm, the sunspot is stabilised by buoyancy forces. The spot is least stable at a depth of about 3\,Mm because of the fluting instability. In deeper layers, the flux tube is marginally unstable. The stability of the sunspot at the surface affects the behaviour of the field lines in deeper layers by magnetic tension. Therefore the fluting instability is damped at depths of about 3\,Mm, and the decrease in compactness is strongest at a depth of about 8\,Mm. The more vertical orientation of the magnetic field and the longer convective timescale lead to slower evolution of the corrugation process in layers deeper than 10\,Mm.}
{The formation of large intrusions of field-free plasma below the surface destabilises the flux tube of the sunspot. This process is not visible at the surface, where the sunspot is stabilised by buoyancy forces. The onset of sunspot decay occurs in deeper layers, while the sunspot still appears stable in the photosphere. The intrusions eventually lead to the disruption and decay of the sunspot.}
\maketitle
%
\section{Introduction}\label{sec:intro}
The lifetime of sunspots covers a broad range of a few days to a few months \citep{MartinezPillet_2002, LR_Driel_2015}. The observational evolution and decay process of sunspots distinguishes a photometric and a magnetic decay \citep{MartinezPillet_2002}. The origin of the differences in lifetime and the flux removal is still not completely understood. In addition, the onset of the decay process is not clear. In some cases, a sunspot already starts to decay, that is, loose magnetic flux, while it might at the same time gain magnetic flux by coalescence \citep{Solanki_2003}. This means that the process that leads to the decay of a sunspot operates already before the sunspot has fully developed \citep{McIntosh_1981}. The opaqueness of the photosphere inhibits direct observations of processes below the solar surface. This limits our ability of fully understanding sunspot evolution from direct observations. Numerical models are used to obtain a more complete image of sunspots.
A variety of different models, especially static ones, has been presented over the years \cite[for an overview of different (static) sunspot models, we refer to][]{Jahn_1997}. The basic assumption for a magnetohydrostatic model is that the sunspot is a monolithic, single magnetic flux tube in and below the photosphere \citep{Schlueter_1958}. For a better agreement of the modelled sunspots with observations, for example, a sharp boundary between the sunspot and the surroundings in intensity, \citet{Simon_1970} implemented a current sheet at the boundary of the flux tube. This current sheet, also called magnetopause, balances the internal pressure with the pressure in the surroundings. The discontinuity of the pressure across the current sheet is correlated to a jump in the field strength \citep{Jahn_1992, Jahn_1994}.\par
\begin{figure}[!t]
\centering 
\includegraphics[width=0.49\textwidth,height=\textheight,keepaspectratio]{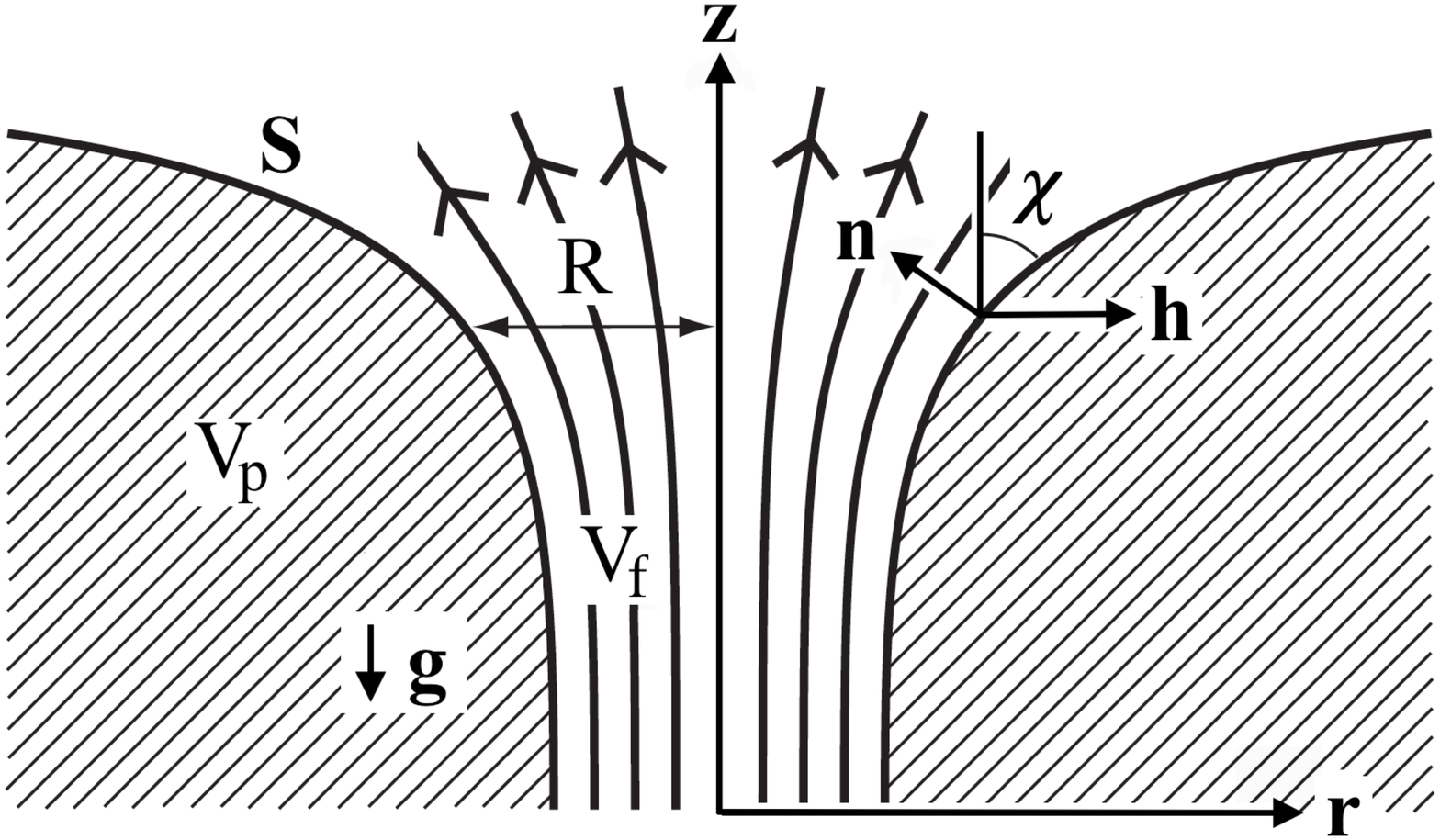}
\caption{Geometry of a magnetic flux tube (V$_\text{f}$) embedded in field-free plasma (V$_\text{p}$) \citep[image adapted from][]{Meyer_1977}.}
\label{fig:model}
\end{figure}

To study the stability of sunspots in the convection zone, \citet{Meyer_1977} adapted a magnetohydrostatic model. They embedded a flux tube with an axisymmetric meridional field $(B_r,0,B_z)$ in the non-magnetised plasma of the upper convection zone. For the magnetic flux tube to be in equilibrium with the surrounding plasma, they assumed the two regions to be separated by a surface S (see Fig.\,\ref{fig:model}). This surface is a current sheet that causes the abrupt drop of the magnetic field and thereby horizontally balances the pressure of the flux tube, $p_\text{f}$, and of the surrounding plasma, $p_\text{p}$. \citet{Meyer_1977} assumed that instabilities of the surroundings, for instance, convection, do not affect the stability of this system. In addition, the plasma in the tube was stably stratified and magnetohydrodynamically stable. Then any perturbation leading to an instability depends on the equilibrium balance at the surface, S. To describe the stability, they took the decreasing pressure difference ($p_\text{p}-p_\text{f}$) with height into account. In addition, the fanning-out of the field with height, which causes a change in inclination angle, $\chi$, of the surface S to the vertical (see Fig.\,\ref{fig:model}) and a change in effect of the gravitational acceleration, ${\bf g}$, onto the magnetised plasma. Finally, the curving of the field causes magnetic tension forces, and the criterion for stability is written as
\begin{equation}\label{eq:meyer_stability}(\rho_{\text{qs}}-\rho_{\text{s}})\cdot g\cdot\text{sin}\chi-\frac{B^2}{4\pi R_c}>0,\end{equation}
with $R_c$ the radius of curvature of the surface S. For a more detailed derivation of the criterion, we refer to \citet{Meyer_1977} and \citet{Strecker_2020}. Based on this inequality, a magnetic flux tube in the upper layers of the convection zone has to balance two main effects to remain stable: (1) The plasma of the magnetic flux tube is less dense than the surrounding plasma, therefore the spot floats on the quiet Sun. (2) The concave curvature of the magnetic field lines has a destabilising effect and causes the tube to become vulnerable to interchange instability.\par
%
Interchange instability is a well-known process in which a cylindrical magnetic flux tube that is perturbed at its outer boundary is deformed. Bundles of field lines are transferred into the non-magnetised neighbourhood. On exchange, regions that were originally occupied by magnetic field lines are occupied by non-magnetised plasma. The individual volumes remain unchanged in this interchange process. The cylindrical flux tube obtains a rippled interface, similar to a fluted column. Therefore the process is also called fluting instability. The direction of the curving of the field lines towards the non-magnetised plasma affects the stability. If the curvature is concave, the rippled interface increases the cross-section and hence destabilises the flux tube further. A detailed explanation of interchange instability can be found in \citet{Priest_2014}.\par
\citet{Schuessler_1984} developed the model of \citet{Meyer_1977} further to study the effect of velocity fields on the instability of flux tubes. He found that converging convective flow cells surrounding the flux tube and flows along the field lines within the flux tube do not have a stabilising effect. Instead, their contribution can be destabilising. Only flows with an azimuthal component around the magnetic flux tube have the ability to stabilise.
\citet{Parker_1979} made use of the stability criterion of \citet{Meyer_1977}. At a sufficiently large depth below the surface, field lines are able to separate from the spot through interchange instability. Parker suggested that a sunspot is not a monolithic, single flux tube below the solar surface. Instead, at about 1Mm below the photosphere, the monolithic structure divides into several smaller flux bundles. Buoyancy and convective downdrafts allow these bundles to bunch together towards the photosphere, and stabilise them against interchange processes \citep{Parker_1979, Jahn_1992}. This model is often referred to as the spaghetti or jelly fish model.\par
Studies performed by means of the photometric evolution of the sunspot take advantage of the change in area of the sunspot in the photosphere in time. A variety of studies have been performed, resulting in different decay processes and scenarios. A linear decay law was found by \citet{Bumba_1963} for recurrent sunspots and was found to fit the distribution in later studies. \citet{MartinezPillet_1993} proposed a parabolic decay rate, with the area being a quadratic function in time. However, they noted that it is difficult to distinguish linear or quadratic decay rates based on the observations. \citet{Petrovay_1997} also proposed a parabolic decay law for their observed sunspots. The determination of a decay law is important to understand the decay process of the sunspots. A linear decay implies that the loss of flux occurs everywhere within the sunspot through turbulent diffusion of the magnetic field \citep[e.g.,][]{Meyer_1974, Solanki_2003}. Instead, a parabolic decay law could be caused by erosion of the sunspot at its boundary, for instance, by supergranular motions \citep{Simon_1964} or turbulent motions on a granular-size scale \citep{Petrovay_1997b}. These models shift the question of an area change during sunspot decay to possible scenarios regarding the loss of magnetic flux from the sunspots. A loss of magnetic flux across the whole area of the sunspot is supported by the appearance of moving magnetic features (MMFs). These appear in the non-magnetised moat region surrounding a sunspot and were first reported by \citet{Sheeley_1969}. \citet{Harvey_1973} were the first to propose that the magnetic patches pull off magnetic flux from the sunspot while moving away. However, later investigations found imbalances between the sunspot magnetic flux loss and the magnetic flux transported by MMFs \citep[e.g.,][]{MartinezPillet_2002, Kubo_2007}. Different types of MMFs were proposed to exist, and were classified according to their polarity \citep{Shine_2000}. Unipolar features of both polarities are observed as well as bipolar features. \citet{MartinezPillet_2002} proposed that only the unipolar features with the same polarity as the sunspot are related to the loss of flux from the sunspot, that is, sunspot decay. Unipolar features with opposite polarity as the sunspot and bipolar features might be extensions of penumbral filaments. Such a scenario was first proposed by \citet{Zhang_1992}. \citet{Thomas_2002} described such a scenario in further detail, with the bipolar features showing a sea-serpent structure. \citet{Kubo_2007}, \citet{Kubo_2012}, and \citet{Verma_2012} also found that the total magnetic flux of the MMFs is far higher than the flux loss of the sunspot. Therefore the polarity of the MMFs has to be considered to determine their relation to sunspot decay. In addition, \citet{Rempel_2015} found in simulations that the emergence and submergence of horizontal field in the moat region also contributes substantially to the flux budget of a sunspot. Specifically, these contributions tend to offset the flux transport due to MMFs.
This might be an explanation for observations of MMFs in the surroundings of stable sunspots that last for several months \citep{LR_Driel_2015}. The relation of MMFs to the decay process of sunspots is still unclear.\par
The knowledge that the flux tube of a sunspot evolves below the photosphere might lead to insights into the decay process. However, the models described above are static. The advent of new computers with more power and larger computational domains enabled setting up realistic magnetohydrodynamic (MHD) simulations in the past years. In this study we make use of a 3D MHD simulation of a sunspot to study the evolution of a sunspot in the upper convection zone and in the photosphere. In Sec.\,\ref{sec:data} we describe the setup and characteristics of the simulation domain that hosts the sunspot. In addition, we determine the flux tube boundary. We analyse the flux tube in Sec.\,\ref{sec:compactness} by studying the evolution of its structure, and by performing a linear stability analysis in Sec.\,\ref{sec:stability}. In both sections we describe the respective methods and results separately. The results of both studies are combined and discussed in Sec.\,\ref{sec:disc}. Finally, a picture of the stability of the flux tube in the upper convection zone and photosphere and a decay scenario of sunspots is presented in Sec.\,\ref{sec:conc}.
\section{Data and methods}\label{sec:data}
We analysed a sunspot in a 3D MHD simulation. It was calculated with the MURaM code by \citet{Rempel_2015}. He calculated two different simulations. We used the simulation that is calculated with a modified numerical diffusivity. For a more detailed description of the calculation and the specification, we refer the reader to Section\,2 of \citet{Rempel_2015}.\par
The simulation domain had a size of 98.308\,$\times$\,98.308\,$\times$\,18.432\,Mm\textsuperscript{3} and was computed on a grid with a 48\,$\times$\,48\,$\times$\,24\,km\textsuperscript{3} cell size. The analysed data set had an increased grid cell size of 96\,km in horizontal direction and a grid spacing of 48\,km in vertical direction. Thus, the analysed 3D domain was composed of 1024\,$\times$\,1024 grid cells horizontal and 384 grid cells vertical. The data show the sunspot with an average diameter of 30\,Mm, the surrounding moat flow with an average extension of 10\,Mm at the surface, and the surrounding quiet Sun. The vertical extension of the simulation domain enabled us to cover sub-photospheric regions with their own characteristics, such as convective motions. The simulation used the setup of \citet{Rempel_2012}, in which a more horizontal field is imposed through the top boundary in order to create an extended penumbra.\par
The complete simulation covered a duration of 100\,hours. In this analysis we considered a section of 30\,hours. This section covers the time range from $t_{\text{sim}}=50$\,h until $t_{\text{sim}}=79.75$\,h, with a cadence of 15\,minutes. For the analysis we used the 3D data of the temperature, $T$, pressure, $p$, density, $\rho$, the three components of the magnetic field ($B_x$, $B_y$, $B_z$), and the velocity ($v_x$, $v_y$, $v_z$). In addition, 2D maps of the magnetic field strength and intensity for selected layers of constant optical depth, $\tau$, were used.\par
The vertical extension of the simulation domain of 18.432\,Mm covers the photosphere and uppermost part of the convection zone. The two regions are attempted to be separated by a solar surface. To define a constant layer within the gridded simulation domain and define it as the solar surface, we determined an average $\tau$\,=\,1 layer. To do this, the optical depth along vertical rays was calculated from the temperature and pressure of each grid cell using the tabulated Rosseland mean opacity, which was also used to advance the grey simulation in the first place. The Rosseland opacity was computed from \citet{Kurucz_1993} opacity tables. The $z$-layer, in the following termed solar surface, is defined as the mean $z$-position where $\tau$\,=\,1. A restriction was made by only taking into account cells outside the sunspot region, that is, the moat and quiet-Sun region. This layer was defined as $z$\,=\,0\,Mm. Thus, the convection zone extends down to $z$\,=\,-17.71\,Mm with negative $z$-values. Regions in the photosphere have positive $z$-values with an extension up to 0.720\,Mm above the solar surface.\par
In this analysis the evolution of the magnetic flux tube of the sunspot is studied from the solar surface downward through the convective region. In this region, we focus on defined depths in steps of $\Delta z$\,=\,0.625\,Mm from $z$\,=\,0\,Mm down to $z$\,=\,-15\,Mm.
\subsection{Convective timescale}\label{sec:convtime}
The evolution of the flux tube of the sunspot is affected by the surrounding regions. We studied the time evolution of the flux tube at different depths in time. Therefore the changing timescales of convective motion within the convection zone had to be considered. In the following we determine the convective timescale for different depths of the simulation domain using the mixing length theory \citep[see e.g.][]{Stix_2002}. In addition, we compare the obtained timescale with the corresponding values of the standard solar model.\par
\begin{figure}[!t]
\centering 
\includegraphics[width=0.49\textwidth,height=\textheight,keepaspectratio]{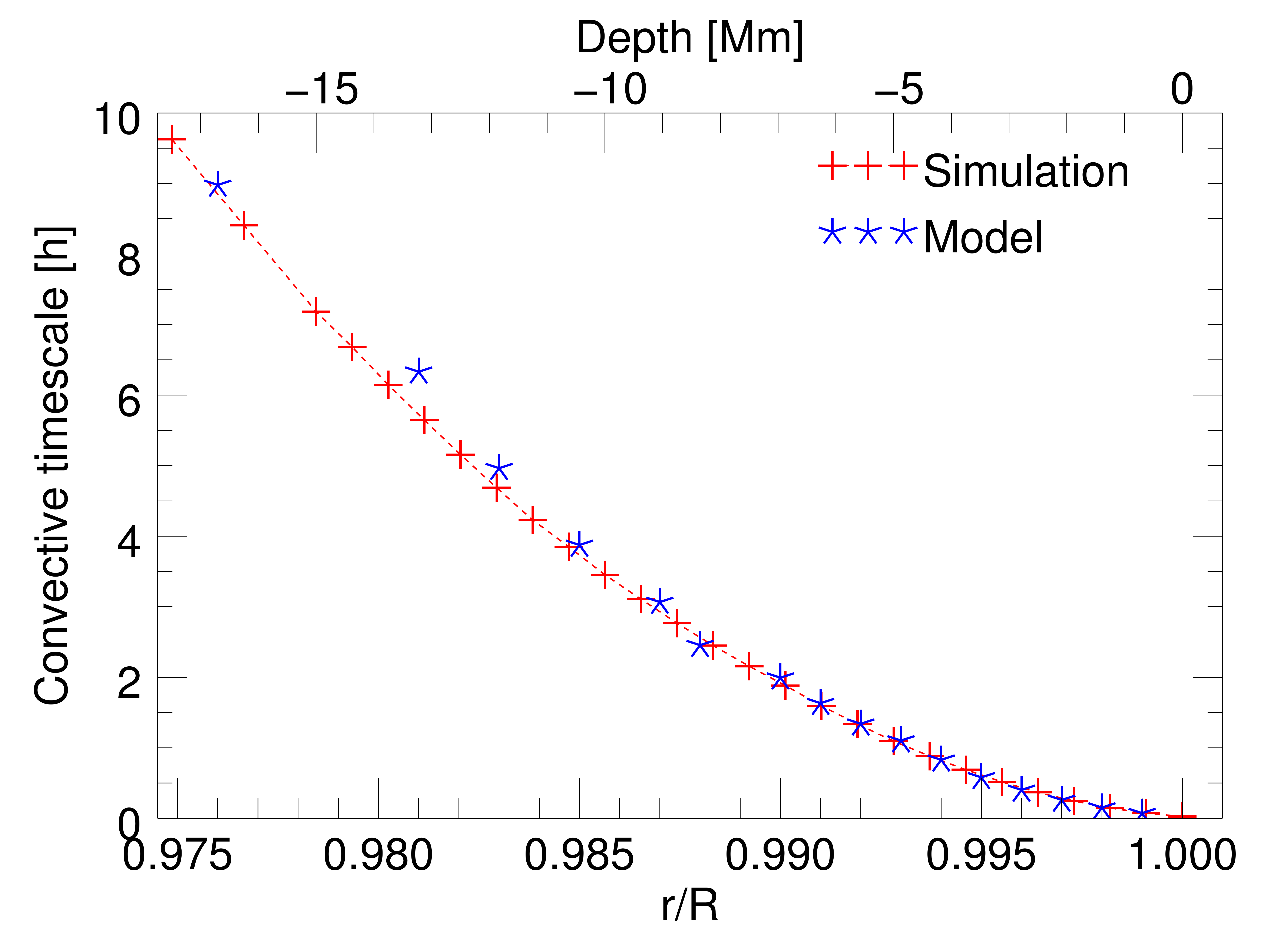}
\caption{Convective timescale $t_\text{conv}$ in the upper part of the convection zone from the MHD simulation (red crosses) and a standard solar model (blue stars) \citep[values are taken from][]{Stix_2002}.}
\label{fig:convtime}
\end{figure}

Based on the mixing length theory, we can describe the evolution of convective motions with the convective timescale, $t_\text{conv}$. It is calculated from the mixing length, $l_M$, and the convective velocity, $v$, as
\begin{equation}t_\text{conv}=\frac{l_m}{v}=\frac{\alpha H_p}{v}.\end{equation}
The mixing length can be approximated by $l_M=\alpha H_p$ with the mixing length parameter, $\alpha$, which is of order unity, and the pressure scale height, $H_p=p/(\rho g)$. Within the mixing length theory, the convective velocity, $v$, is described as the average velocity of a rising bubble through a defined layer. To determine the convective timescale within the simulation domain, the thermodynamic quantities were determined in steps of $\Delta z$\,=\,0.625\,Mm starting at $z$\,=\,0\,Mm down to $z$\,=\,-15\,Mm, and additionally, the depths $z$\,=\,-16.25\,Mm and $z$\,=\,-17.5\,Mm were considered. Only the region outside the sunspot was used to determine the thermodynamic properties of the simulated convection zone. There, in quiet-Sun convection, $\alpha=1$ is a good approximation. For the velocity, the absolute value of the vertical component $v_z$ was used. For each depth position, we first determined the mean over nine layers (four layers above and four below the respective layer) and finally the mean over all time steps. The resulting local timescale for the different depths in the convective region of the simulation domain are shown in Fig.\,\ref{fig:convtime} as red crosses. The increase in timescale with depth represents a slower evolution of perturbations with deeper layers. Blue stars in Fig.\,\ref{fig:convtime} show the convective timescale calculated with values based on the standard solar model as provided by \citet[][]{Stix_2002}. A comparison of the values from the simulation and the standard solar model shows that the thermodynamic quantities of the two agree well.
\subsection{Determination of the boundary of the sunspot flux tube}\label{sec:bcrit}
The sunspot region needs to be clearly defined when the structure of the magnetic flux tube of the sunspot is to be analysed. To do this, we used the magnetic field strength, $B=\sqrt{B_x^2+B_y^2+B_z^2}$. At each of the 25 depth positions from $z$\,=\,0\,Mm to $z$\,=\,-15\,Mm, an individual value of the magnetic field strength was determined as a contour value to define the boundary of the magnetic flux tube.\par
We applied two different methods to determine the contour values. The first method is an estimate of the boundary by eye. The values $B_{\text{eye}}(z)$ were chosen to obtain the best match between the contours and the magnetic flux tube at the individual depths. To this end, contours for different $B$ values were visualised on horizontal $B$ maps, clipped to $B\le6$\,kG, for arbitrary time steps. We then chose the contour that best represented the visual contrast between the magnetic flux tube and the non-magnetic surroundings. The obtained values, determined in steps of $\Delta z$\,=\,1.25\,Mm, are shown in Fig.\,\ref{fig:bcrit} as blue stars and are listed in Table\,\ref{tab:bcrit} in Appendix\,\ref{app:bcrit}. This manual method is neither consistent nor applicable to the large number of contours.\par
As second method we applied an automatic procedure to determine the threshold $B_{\rm c}(z)$. The maximum magnetic field strength $B^{\text{max}}(z,t)$ was determined for each grid layer and each time step. $B^{\text{max}}(z,t)$ was boxcar-smoothed with a width of 240\,km in vertical direction and a width of 75\,minutes in time around a defined depth position in steps of $\Delta z$\,=\,0.625\,Mm. From the smoothed values, averages, $\langle B^{\text{max}}_{\text{smooth}}\rangle_{t, z\pm5}$, over all times and ten adjacent depth layers were calculated. These values were divided by $2\text{e}$ to yield the threshold $B_{\rm c}(z)$,
\begin{equation}B_\text{c}(z)=\frac{\langle B^{\text{max}}_{\text{smooth}}\rangle_{t, z\pm5}}{2e}.\end{equation}
We made tests for multiples of $e$ in visualising the different contours on horizontal $B$ maps for three different times of 0\,h, 15\,h, and 27.75\,h and for depths in steps of $\Delta z=1.25$\,Mm. The best representation of the boundary of the magnetic flux tube throughout the different depth positions was found for a fraction of $2e$ of $\langle B^{\text{max}}_{\text{smooth}}\rangle_{t, z\pm5}$. With this choice, the contours from $B_{\text{c}}$ and $B_{\text{eye}}$ match fairly well, although they are determined independent of each other, but differences exist in a few cases.\par
The obtained $B_\text{c}(z)$-values are shown as red crosses in Fig.\,\ref{fig:bcrit} overlaid on the temporal averaged values of the maximum magnetic field strength for each vertical grid layer, $B^{\text{m,t}}(z)/2e$ (black dots). The calculated contour values $B_\text{c}(z)$ are given in Table\,\ref{tab:bcrit} in Appendix\,\ref{app:bcrit} as well.\par
%
\begin{figure}[!ht]
\centering 
\includegraphics[width=0.49\textwidth,height=\textheight,keepaspectratio]{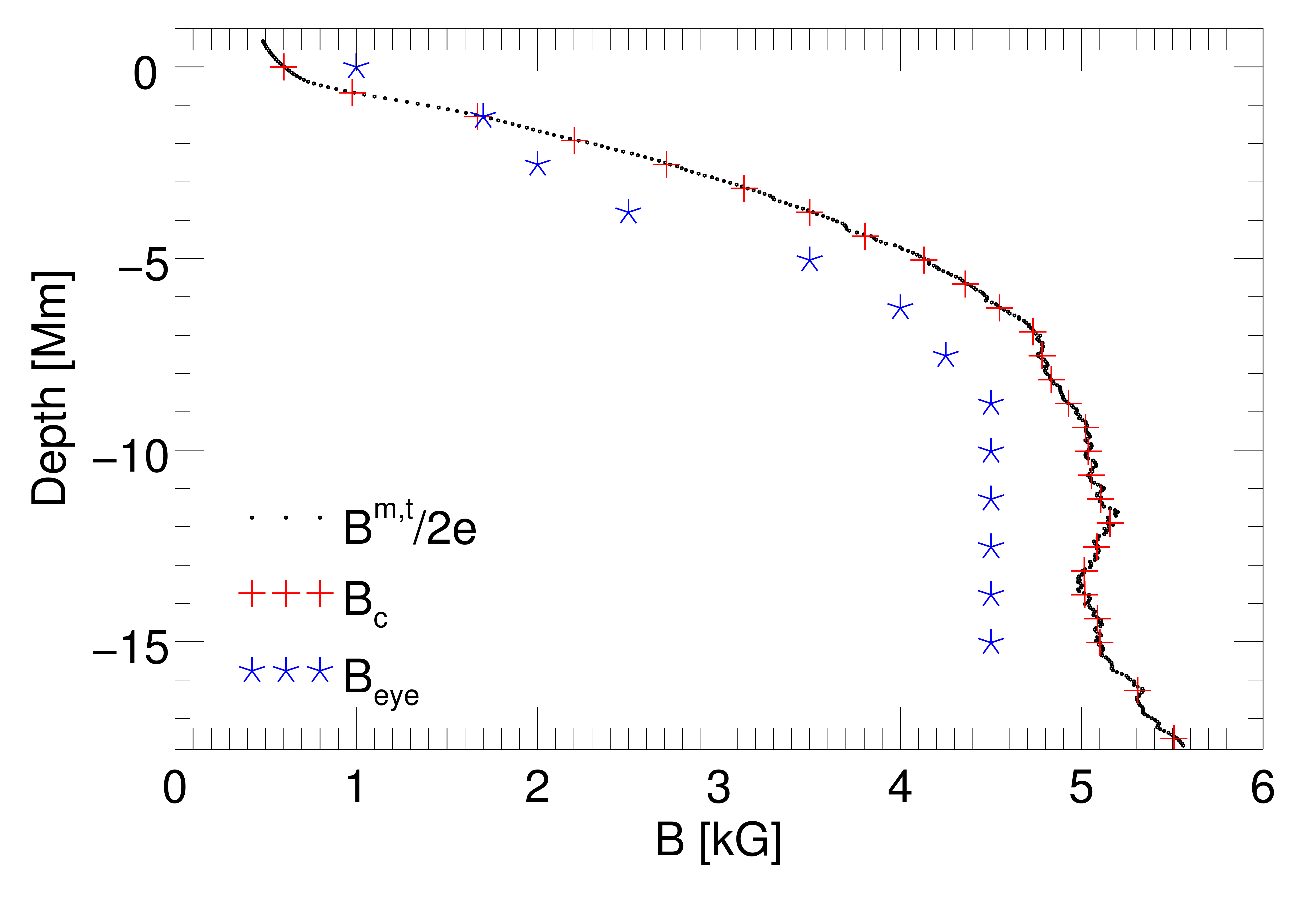}
\caption{Contour values of the magnetic field strength ($B_\text{eye}(z)$ and $B_\text{c}(z)$) used to determine the sunspot boundary by eye (blue) and from the maximum magnetic field strength (red).\label{fig:bcrit}}
\end{figure}
 We computed the area, $A$, and perimeter, that is, the length of the contour, $C$, of the sunspot contours at each depth and time step as follows: We smoothed each map of the magnetic field strength with a width of 5 in $x$- and $y$-\,direction. We applied the values $B_{\text{eye}}(z)$ and $B_{\text{c}}(z)$ to determine the contours.
\section{Compactness of the sunspot}\label{sec:compactness}
\begin{figure}[!th]
\centering 
\includegraphics[width=0.49\textwidth,height=\textheight,keepaspectratio]{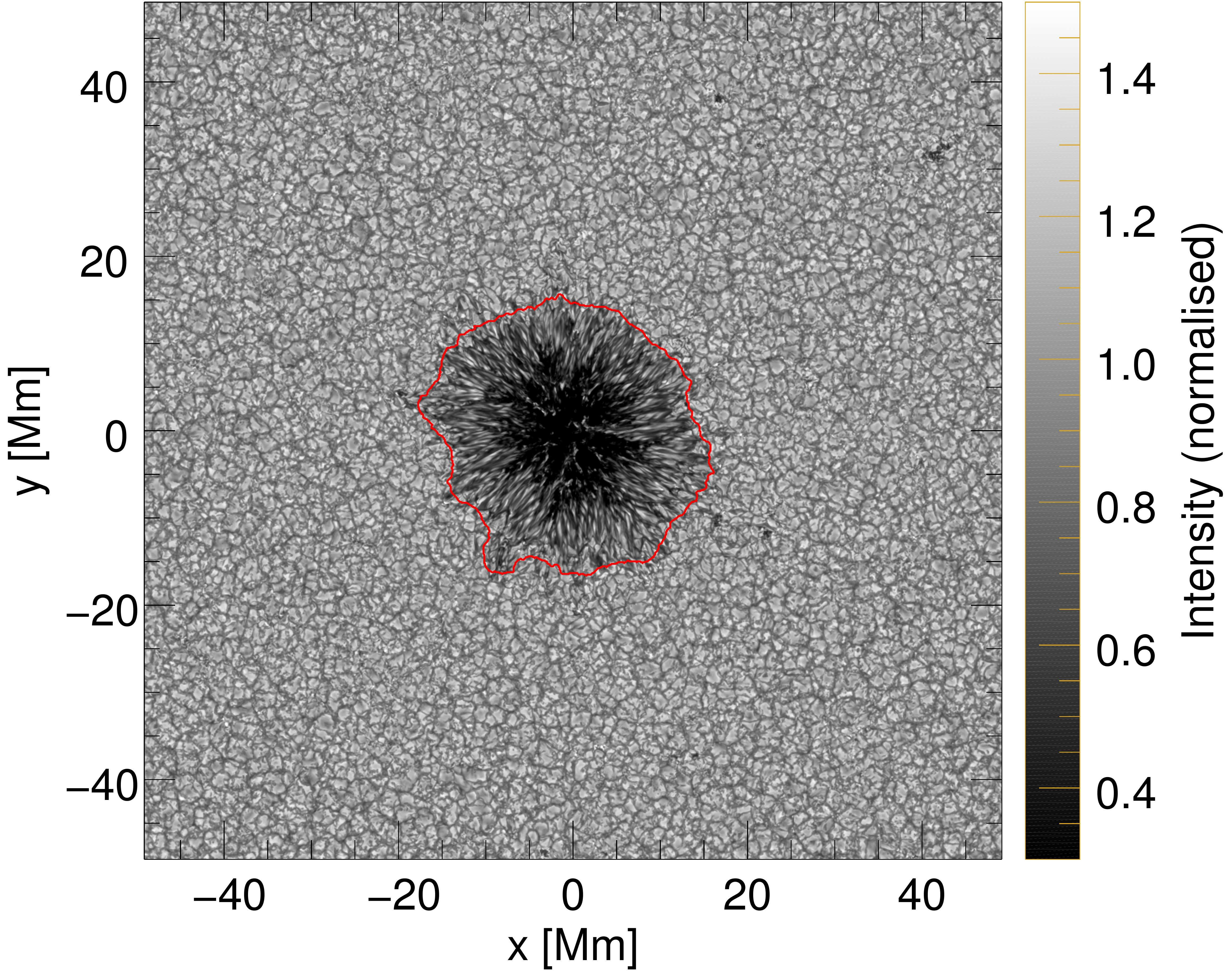}
\caption{Intensity map at $\tau=1$ with a contour line at $I_{\text{c}}=0.9I_{\text{n}}$ indicating the boundary of the sunspot.}
\label{fig:int}
\end{figure}

\begin{figure*}[!hp]
\includegraphics[height=0.96\textheight,width=\textwidth,keepaspectratio]{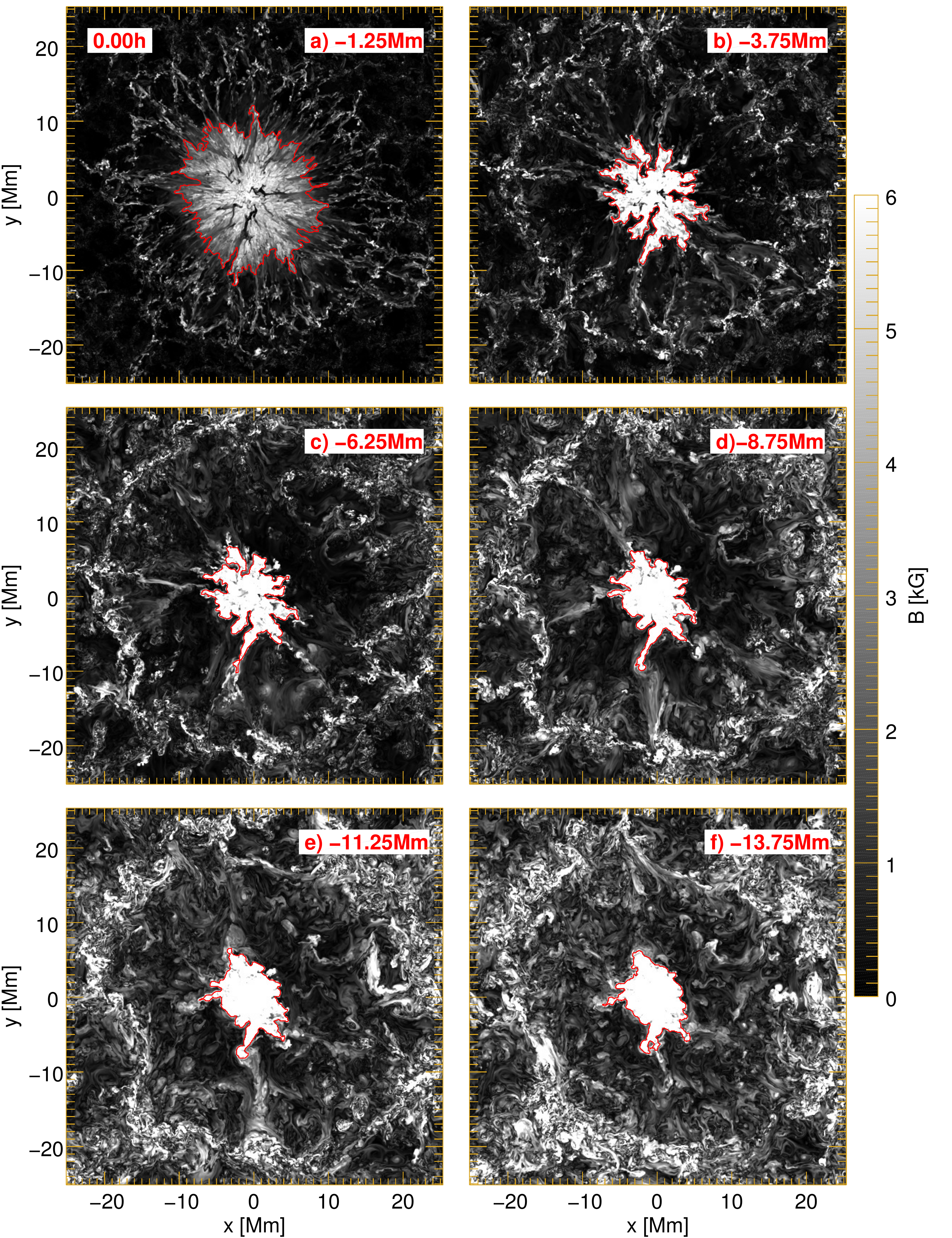}
\caption{Maps of the magnetic field strength at $t=0$\,h, shown at different depths (indicated in red in each panel). The boundary of the sunspot (red line) is defined by depth-dependent values of the magnetic field strength.}
\label{fig:bmapsdepth}
\end{figure*}

Because of the filamentary structure of the penumbra, the sunspot is not perfectly circular, as shown in the intensity map in Fig.\,\ref{fig:int}, which corresponds to $t$\,=\,0\,h. A contour, outlining the boundary of the sunspot, is drawn at an intensity level $I_{\text{c}}=0.9I_{\text{n}}$, with the normalised intensity, $I_{\text{n}}=I/\bar{I}_{\text{qs}}$, where $\bar{I}_{\text{qs}}$ denotes the mean intensity of the quiet Sun. A similar structure can be found in maps of the magnetic field strength at 1.25\,Mm below the surface. This is shown for the same time step in Fig.\,\ref{fig:bmapsdepth}\,a. The outer boundary of the flux tube, shown in red, is determined with the automated procedure as described in Sec.\,\ref{sec:bcrit}. At this depth, the flux tube is embedded in a mesh-like structure that covers the moat. This structure terminates at the network boundary. This network boundary is seen throughout all depth layers, as displayed in Fig.\,\ref{fig:bmapsdepth}\,b\,-\,f, meaning that the sunspot is embedded in the centre of a moat cell \citep[see also][]{Rempel_2015}. We note that in Fig.\,\ref{fig:bmapsdepth} the field of view is reduced to focus on the sunspot and the moat.\par
Field-free regions penetrate the flux tube. In deeper regions, the average diameter of the flux tube decreases and the field strength increases. In addition, the boundary of the tube becomes more ragged, and the indentations at the boundary penetrate deeper into the inner part of the flux tube. This is most prominent at depths of $z$\,=\,-3.75\,Mm and $z$\,=\,-6.25\,Mm, as shown in Fig.\,\ref{fig:bmapsdepth}\,b and c, respectively. The flux tube is more compact in deeper regions, that is, the outer boundary becomes smoother (see Figs.\,\ref{fig:bmapsdepth}\,d\,-\,f).\par
%
\begin{figure*}[!hp]
\centering 
\includegraphics[height=0.96\textheight,keepaspectratio]{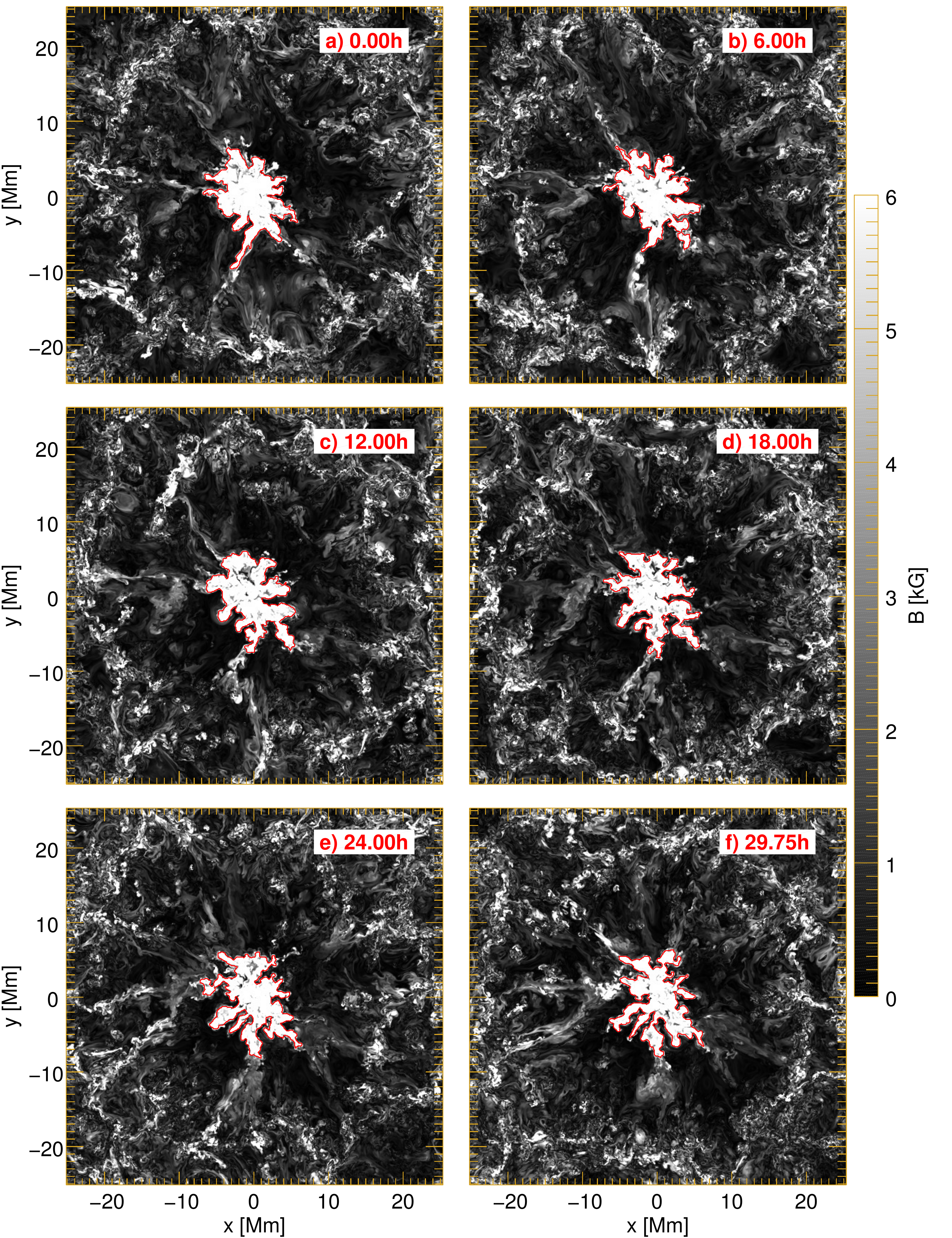}
\caption{Maps of the magnetic field strength at $z$\,=\,-7.5\,Mm at different times (indicated in red in each panel). The boundary of the sunspot (red line) is defined by the contour value $B_\text{c}=4781$\,G.}
\label{fig:bmaps75Mm}
\end{figure*}

In time, the ragged structure of the flux tube becomes more pronounced. This is shown in Fig.\,\ref{fig:bmaps75Mm} for a depth of $z$\,=\,-7.5\,Mm with a cadence of 6\,hours between the snapshots, covering the 30 hours of the analysed data. An animation with a cadence of 15\,minutes for selected depths is provided as additional material\footnote{The movies associated to Fig.\,\ref{fig:bmaps75Mm} are available at \url{http://www.leibniz-kis.de/fileadmin/user_upload/kis/media/promotionen/2020_strecker/BMaps.mp4}}. 
At the beginning of the analysis, at $t$\,=\,0\,h (see Fig.\,\ref{fig:bmaps75Mm}\,a) the inner part of the flux tube is mainly undisturbed. Within 6\,hours, regions of weaker field appear in the innermost part of the flux tube while the outer structure becomes more ragged, as is shown in Fig.\,\ref{fig:bmaps75Mm}\,b. The increasing raggedness causes a degradation of the flux tube. This degradation process continues in time, as Figs.\,\ref{fig:bmaps75Mm}\,c\,-\,e show. At 29.75\,h, the roundish structure of the flux tube has completely vanished. The degradation process takes place in all regions deeper than 1\,Mm below the surface (see the animation provided as additional material).
\subsection{Method}\label{sec:comp_method}
For a quantitative description of the degradation process, we define the compactness, $K$, as the ratio of the circumference, $2\sqrt{\pi A}$, that corresponds to a circle with the area $A$, and the actual perimeter, $C$, around the magnetised area $A$, that is, the length of the contour with area $A$,
\begin{equation}K=2\sqrt{\pi}\frac{\sqrt{A}}{C}.\label{eq:compactness}\end{equation}
This enables a comparison of different depths as the average diameter of the flux tube decreases with depth. Thus, a circular flux tube has $K=1$. The compactness $K(z,t)$ was calculated for all time steps and all selected depths.\par
For an increasing raggedness, as discussed in context of Fig.\,\ref{fig:bmaps75Mm}, the length of the contour increases. This causes a decrease in the compactness $K$ in time, as described by Eq.\,\eqref{eq:compactness}.
\subsection{Results}\label{sec:comp_results}
\begin{figure*}[!hp]
\centering 
\includegraphics[height=0.94\textheight,width=\textwidth,keepaspectratio]{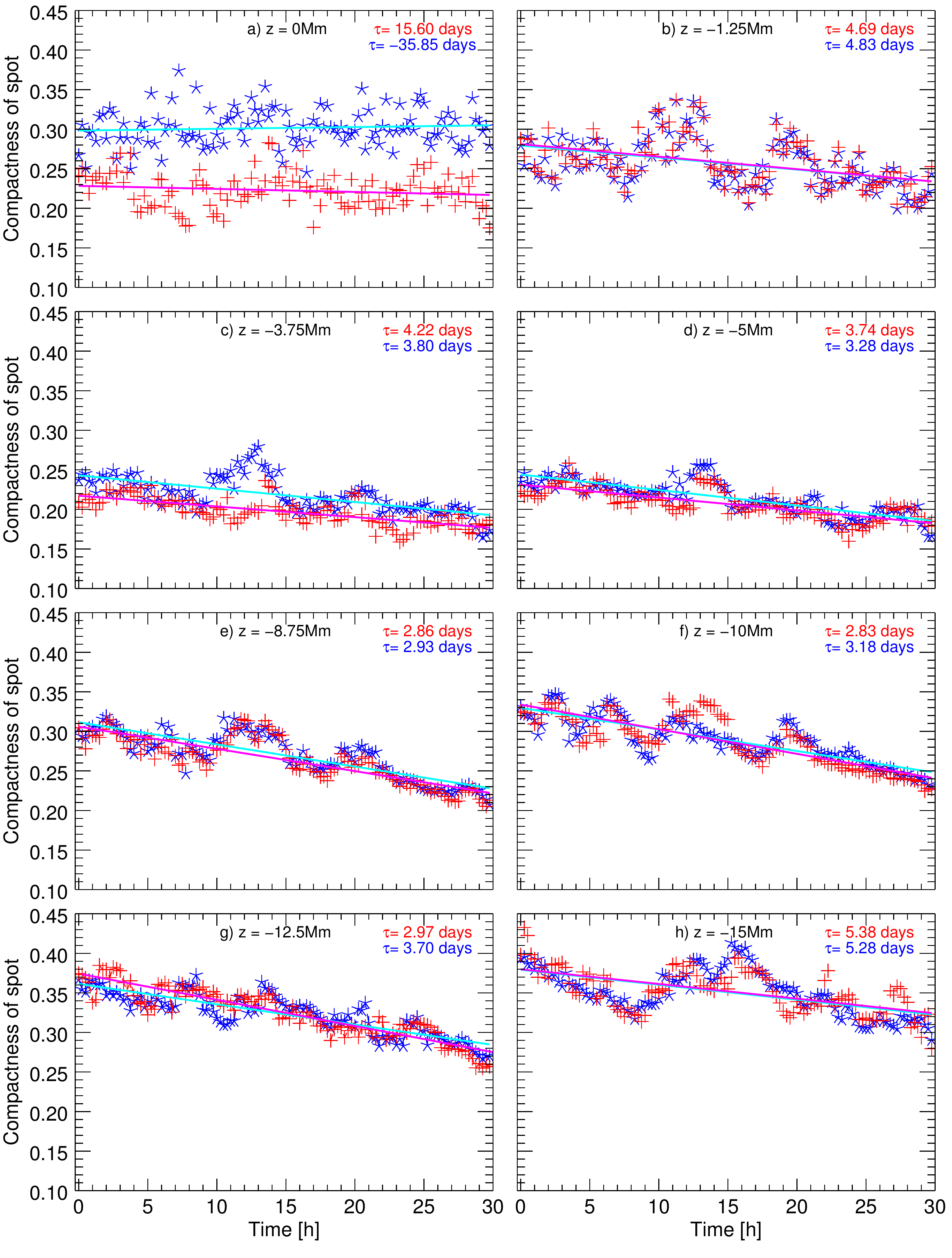}
\caption{Compactness of the sunspot (normalised area-perimeter ratio) in time, shown at different depths. The boundary of the sunspot is determined with depth-dependent contour values $B_{\text{c}}(z)$ (red crosses) and $B_{\text{eye}}(z)$ (blue stars). Degradation times are calculated from linear fits (magenta and cyan lines).\label{fig:compactness}}
\end{figure*}

The compactness in time for eight selected depths is shown in Fig.\,\ref{fig:compactness}. The compactness evolution for all other analysed depths is shown in Fig.\,\ref{fig:compactness_app1} and \ref{fig:compactness_app2} in Appendix\,\ref{app:compactness}. The determination of the sunspot boundary might affect the results. Therefore we analysed a sunspot boundary with both contour values, $B_{\text{c}}$ and $B_{\text{eye}}$ (see Sec.\,\ref{sec:bcrit} and Fig.\,\ref{fig:bcrit}, red crosses and blue stars, respectively). However, the analysis with $B_{\text{eye}}$ was only performed for depths in steps of $\Delta z$\,=\,1.25\,Mm. Therefore no values are available at a depth of $z$\,=\,-1.875\,Mm, for instance. If not explicitly noted otherwise, the following analysis describes the results obtained with $B_{\text{c}}(z)$.\par
%
At the surface, the compactness of the flux tube is almost constant (see Fig.\,\ref{fig:compactness}\,a). The scattering of the values is caused by the arbitrary change in filamentary structure of the penumbra. At $z$\,=\,-0.625\,Mm the compactness is much smaller than at other depths (see Fig.\,\ref{fig:compactness_app1}\,a). Here, the area of the sunspot is overestimated. The contour value $B_{\text{c}}$(-0.625\,Mm) is not able to reproduce the sunspot boundary. It includes small magnetic structures of the surrounding region. Thus, a higher value for the area leads to a smaller compactness (see Eq.\,\eqref{eq:compactness}). In conclusion, these results and the visual impression of the sunspot in maps of the magnetic field strength lead to the picture of an overall stable sunspot above $z$\,=\,-1\,Mm. This agrees with the evolution of the sunspots magnetic flux at the surface, which is almost stationary, according to \citet{Rempel_2015}. Below $z$\,=\,-1\,Mm, the compactness decreases, as we show in Fig.\,\ref{fig:compactness}\,b\,-\,h and Fig.\,\ref{fig:compactness_app1}\,b\,-\,h and \ref{fig:compactness_app2} in Appendix\,\ref{app:compactness}.\par
The initial compactness differs for different depths. A minimum compactness is found 3.75\,Mm below the surface (see Fig.\,\ref{fig:compactness}\,c). Towards higher and deeper regions the compactness is larger. The maximum compactness is found at the lowest analysed depth, that is, 15\,Mm below the surface. This means that at the beginning of the analysis, the sunspot structure is ragged most strongly at 3.75\,Mm below the surface. This agrees with the first visual impression obtained from maps of the magnetic field strengths as described above and shown in Fig.\,\ref{fig:bmapsdepth}.\par
A similar compactness is found for an analysis with different contour values, that is, $B_{\text{c}}(z)$ and $B_{\text{eye}}(z)$. The compactness evolutions for the individual depths resemble each other. We can therefore assume that the exact determination of the sunspot boundary does not substantially affect the analysis of the sunspot compactness. The decrease in compactness in time for all depths reproduces the increase in raggedness, as described above. The evolution of the compactness can be approximated by a linear fit (magenta and cyan lines in Fig.\,\ref{fig:compactness}), despite the non-continuous decrease. The gradients of the fits show that the compactness does not decrease at the same rate at different depths.
\paragraph{Degradation time of the sunspot:}\label{sec:comp_decaytime} A decrease in compactness describes an increase in raggedness of the flux tube. A higher raggedness causes an increase in the outer surface of the flux tube. Thus, the contact area of the magnetic flux tube with the non-magnetic surrounding plasma increases. This enables a stronger effect of the surroundings, that is, plasma motions. We approximated the evolution of the compactness by a linear fit (see Fig.\,\ref{fig:compactness}) to calculate a degradation time of the sunspot by the time at which the initial value of the fit has dropped by a factor of $1/e$.\par
\begin{figure}[!ht]
\centering 
\includegraphics[width=0.49\textwidth,height=\textheight,keepaspectratio]{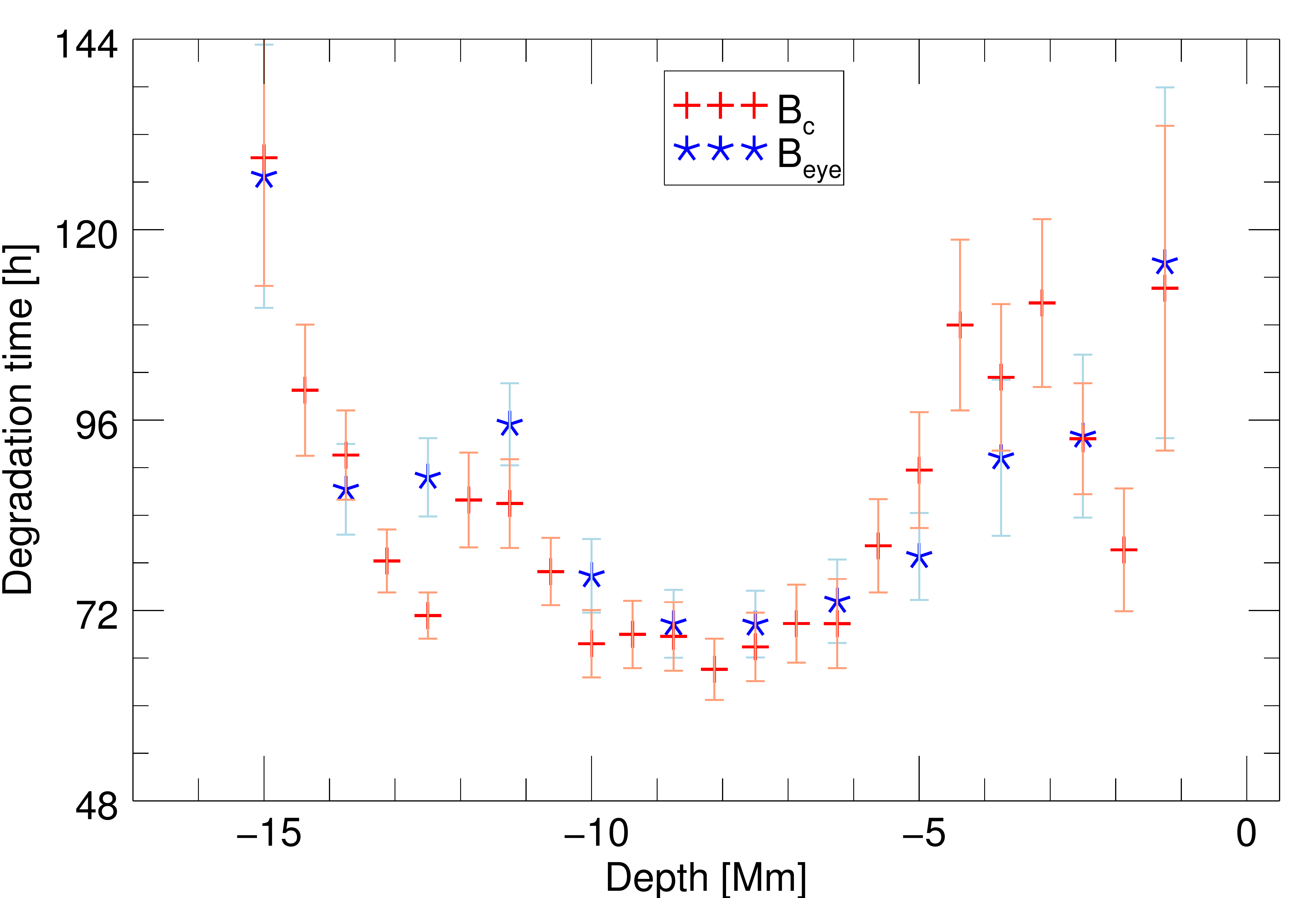}
\caption{Degradation time of the sunspot at different depths from the compactness of the flux tube. The analysis was made with different depth-dependent contour values $B_\text{c}(z)$ and $B_\text{eye}(z)$ to determine the sunspot boundary (see Sec.\,\ref{sec:bcrit}).}
\label{fig:decay_time}
\end{figure}
\begin{figure}[!ht]
\centering 
\includegraphics[width=0.49\textwidth,height=\textheight,keepaspectratio]{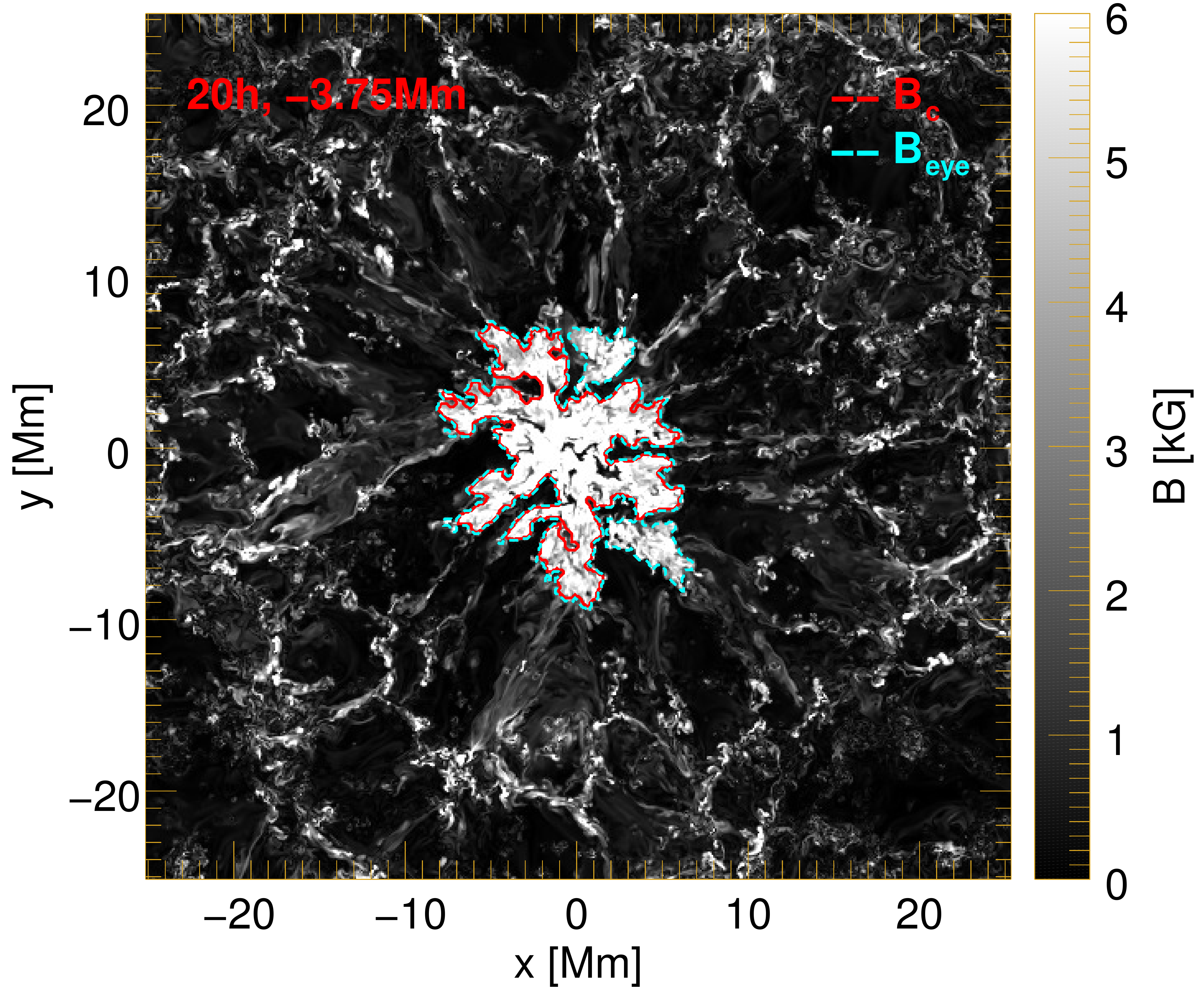}
\caption{Map of the magnetic field strength at $z$\,=\,-3.75\,Mm, at $t$\,=\,20\,h. The outlined contour of the sunspot is determined with $B_{\text{c}}$\,=\,3500\,G (red) and $B_{\text{eye}}$\,=\,2500\,G (cyan).}
\label{fig:contvar}
\end{figure}

Above $z$\,=\,-1\,Mm the compactness is stable, as described and concluded above. Therefore the following analysis focuses on layers deeper than 1\,Mm below the surface. The degradation time is calculated for the same depths as the compactness. We also used the compactness obtained for both methods, that is, a determination of the flux tube with $B_{\text{c}}$ and $B_{\text{eye}}$, to calculate the degradation times. A similar distribution of the degradation times with depth can be found for $B_{\text{c}}$ and $B_{\text{eye}}$ as shown in Fig.\,\ref{fig:decay_time} (red crosses and blue stars, respectively). Error bars show the standard deviation obtained from the linear fits. Discrepancies between the degradation times when applying different contour values for individual depths can be ascribed to the evolution of the field strength within the flux tube. For times $t\ge20$\,h at depths between 3\,Mm and 5\,Mm below the surface, the contours of the flux tube obtained with the different methods are significantly different. Fig.\,\ref{fig:contvar} shows a map of the magnetic field strength at a depth of 3.75\,Mm below the surface at $t$\,=\,20\,h where the contour lines obtained with the two methods differ significantly. This is one of the few cases where the automatic and manual contours do not match: The higher value of $B_{\text{c}}$ (red) excludes two regions that are included when $B_{\text{eye}}$ (blue) is applied. This affects the determined compactness and leads to the deviation of the degradation times for $B_{\text{c}}$ and $B_{\text{eye}}$.\par
%
The structure of the flux tube evolves fastest in regions between 6.25\,Mm and 10\,Mm below the surface. There, we measure the fastest degradation times of less than three\,days, as Fig.\,\ref{fig:decay_time} shows. The slowest decrease in compactness of the flux tube is found at a depth of 15\,Mm with a maximum degradation time of 5.4\,days.
\section{Stability of the sunspot}\label{sec:stability}
To understand the decay process of a sunspot, we took stabilising as well as destabilising effects into account. To do this, we studied the stability of the simulated sunspot in the convection zone.
\subsection{Methods}\label{sec:stab_methods}
The stability of a magnetic flux tube in the upper convection zone is affected by two main effects: The stratification of the thermodynamic properties such as density within the flux tube and in the direct surroundings, and the curvature of the outer surface of the flux tube defining the sunspot. \citet{Meyer_1977} described the stability of a flux tube in the upper convection zone and took both effects into account (see Sec.\,\ref{sec:intro}). We used Eq.\,\eqref{eq:meyer_stability} to study the stability of the flux tube of the simulated sunspot in the convection zone. To do this, we determined the following parameters: (1) The density difference in the flux tube, $\rho_\text{s}$, and the surroundings, $\rho_\text{qs}$, as a function of depth. (2) The inclination angle, $\chi$, of the outer surface of the flux tube towards the vertical and its radius of curvature, $R_\text{c}$. Both parameters are obtained from a surface function that describes the radial extension of the sunspot with depth.
\subsubsection{Density stratification in the upper convection zone}\label{sec:stab_density}
\begin{figure}[!ht]
\centering 
\includegraphics[width=0.49\textwidth,height=\textheight,keepaspectratio]{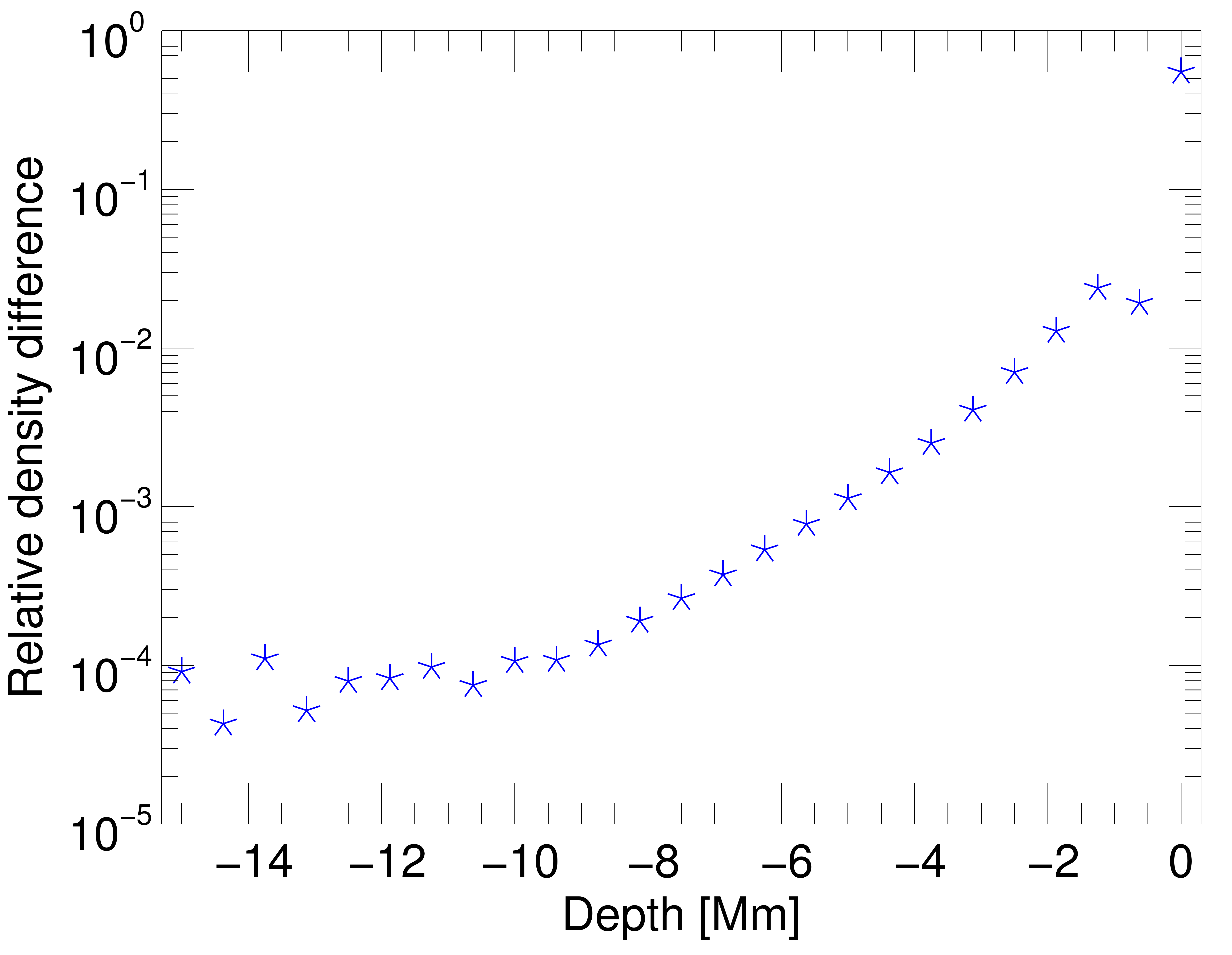}
\caption{Relative density difference between the quiet Sun and the magnetic flux tube, time-averaged, at different depths.}
\label{fig:density_diff}
\end{figure}

From observations we know that temperature, pressure, and density within a sunspot are lower than in the surrounding quiet Sun. However, this only holds for regions close to the surface. In deeper regions, these differences are expected to decrease \citep[e.g. ][]{Jahn_1994, Schlichenmaier_1997}. We determined spatial averages of the density for all defined depths $z$ and all time steps for the areas covered by the flux tube of the sunspot and the quiet Sun. The relative density difference between the two regions, $(\rho_\text{qs}-\rho_\text{s})/\rho_\text{qs}$, decreases with increasing depth, as Fig.\,\ref{fig:density_diff} shows for time-averaged values. Because we averaged the spot density over umbra and penumbra, the resulting density contrast just beneath the surface is affected by the Wilson depression, which leads to very low densities in the umbra. Together with a misinterpretation of the flux tube contour, this leads to the discontinuities at $z>-0.625$\,Mm.
\subsubsection{Surface function of the flux tube}\label{sec:stab_surfacefunc}
\begin{figure}[!ht]
\centering 
\includegraphics[width=0.49\textwidth,height=\textheight,keepaspectratio]{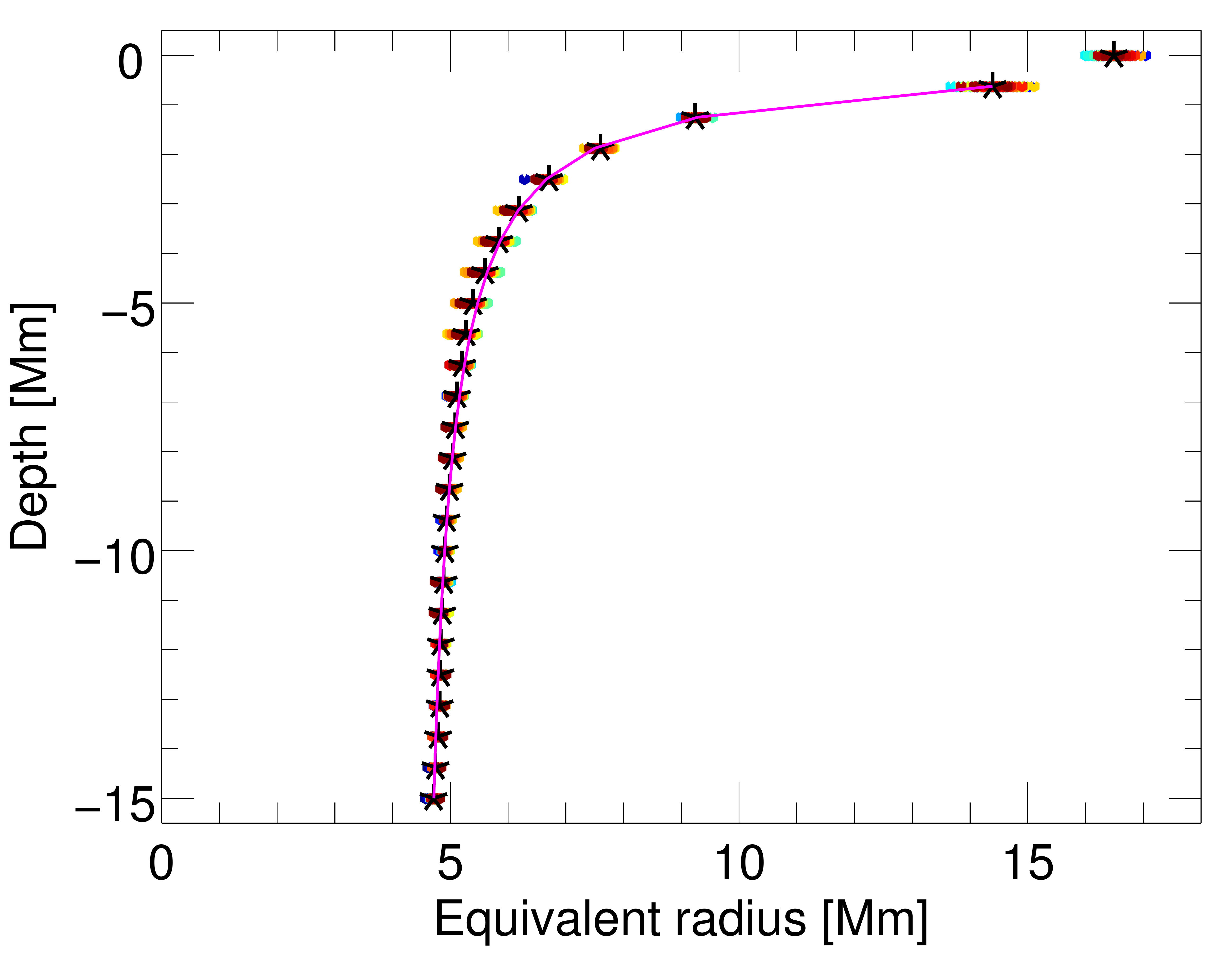}
\caption{Radial extension of a circular flux tube at different depths obtained from the area $A$ of the simulated sunspot. The time-averaged values (black stars) of all 120 time steps (rainbow-coloured dots, red to blue in time) are fitted by an asymptotic function (magenta line).}
\label{fig:surface_func}
\end{figure}

We define an equivalent radius of the flux tube as the radius, $r_{\rm e}=\sqrt{A/\pi}$, of a circle with the same area (see Sec.\,\ref{sec:comp_method}).
The equivalent radius, $r_{\text{e}}(z,t)$, is calculated for all defined depths $z$ and all 120 time steps. These individual values are shown in Fig.\,\ref{fig:surface_func} as rainbow-coloured dots. To approximate the depth dependence, we constructed an asymptotic function,
\begin{equation}r_{\text{e}}(z,t)=\frac{a_{0}z^{2}+a_{1}}{a_{2}+a_{3}z^{3}}+a_{4}\,,\end{equation}
with fitting parameters, $a_{i}$. The calculation was made for all 120 time steps with $r_{\text{e}}(z,t)$ over a depth range from $z$\,=\,\text{-}0.625\,Mm to $z$\,=\,\text{-}15\,Mm. Fig.\,\ref{fig:surface_func} shows the best-fit function (magenta line), obtained from time-averaged equivalent radii (black stars). The obtained parameters for this fit function are $a_{0}=-10.38$\,Mm\textsuperscript{-1}, $a_{1}=-2.74$\,Mm, $a_{2}=-0.22$\,Mm, $a_{3}=1.85$\,Mm\textsuperscript{-2}, and $a_{4}=4.34$\,Mm with $\chi^{2}_\text{gof}=0.002$ as the goodness of the fit (gof).\par
%
Inclination angle $\chi$: The inclination angle $\chi$ is calculated from the derivative of the surface function with
\begin{equation}\chi(z)=\tan^{\text{-1}}(r'_{\text{e}}(z))=\tan^{\text{-1}}\left(\frac{z\cdot(2a_0a_2-a_0a_3z^3-3a_1a_3z)}{(a_2+a_3z^3)^2}\right).\end{equation}
At each depth position $z\le$\,-0.625\,Mm and for each time step the spatial derivative $r'_{\text{e}}(z,t)$ is calculated. A minimum angle of less than 1.7$^{\circ}$ is found at the lower-most depth, that is, 15\,Mm below the surface. The inclination angle slowly increases towards the surface and reaches 13$^{\circ}$ at $z=-5$\,Mm. Then, in the upper 5\,Mm of the convection zone the increase steepens and almost reaches 85$^{\circ}$ at $z$\,=\,-0.625\,Mm.\par
\begin{figure}[!t]
\centering 
\includegraphics[width=0.49\textwidth,height=\textheight,keepaspectratio]{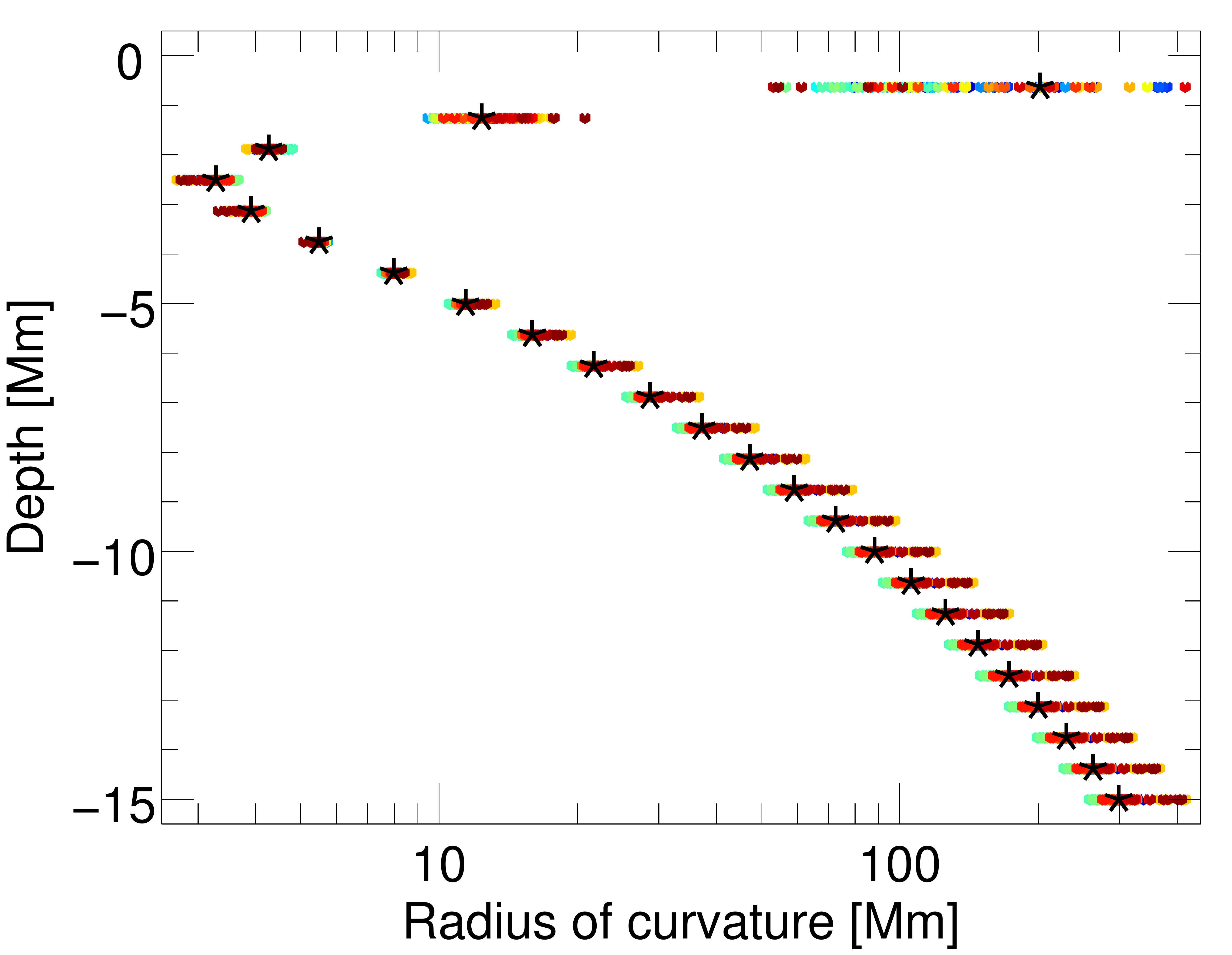}
\caption{Radius of curvature at different depths, time-averaged (black stars) over all 120 time steps (rainbow-coloured dots,
blue to red in time).\label{fig:rad_curv}}
\end{figure}

Radius of curvature $R_\text{c}$: To calculate the radius of curvature, the first and second derivatives, $r'_{\text{e}}(z,t)$ and $r''_{\text{e}}(z,t)$, respectively, of the surface function have to be calculated. It is then determined for each depth position $z\le$\,-0.625\,Mm and time step individually with
\begin{equation}R_\text{c}(z)=\frac{(1+r'_{\text{e}}(z)^2)^{3/2}}{|r''_{\text{e}}(z)|}.\end{equation}
The radius of curvature is largest at the lowest measurement point, that is, 15\,Mm below the surface (see Fig.\,\ref{fig:rad_curv}). The flux tube is almost vertical there, thus, the curvature of the flux tube is smallest. The radius of curvature decreases with decreasing depth and has a minimum at $z$\,=\,-2.5\,Mm. Upward of the minimum, the radius of curvature increases again. Because of the misinterpretation of the sunspot boundary at $z$\,=\,-0.625\,Mm (see Sec.\,\ref{sec:comp_results}), the corresponding $R_{\text c}$ values show a large scatter, but the trend of an upward-increasing curvature radius is obvious.\par
%
Another parameter necessary to apply the stability criterion, that is, Eq.\,\eqref{eq:meyer_stability}, to the simulated sunspot is the magnetic field strength $B(z,t)$. It is obtained as spatial average over the area of our flux tube contours for each time step $t$ and each selected depth $z$.
\subsection{Results}\label{sec:stab_results}
\begin{figure}[!t]
\centering 
\includegraphics[width=0.49\textwidth,height=\textheight,keepaspectratio]{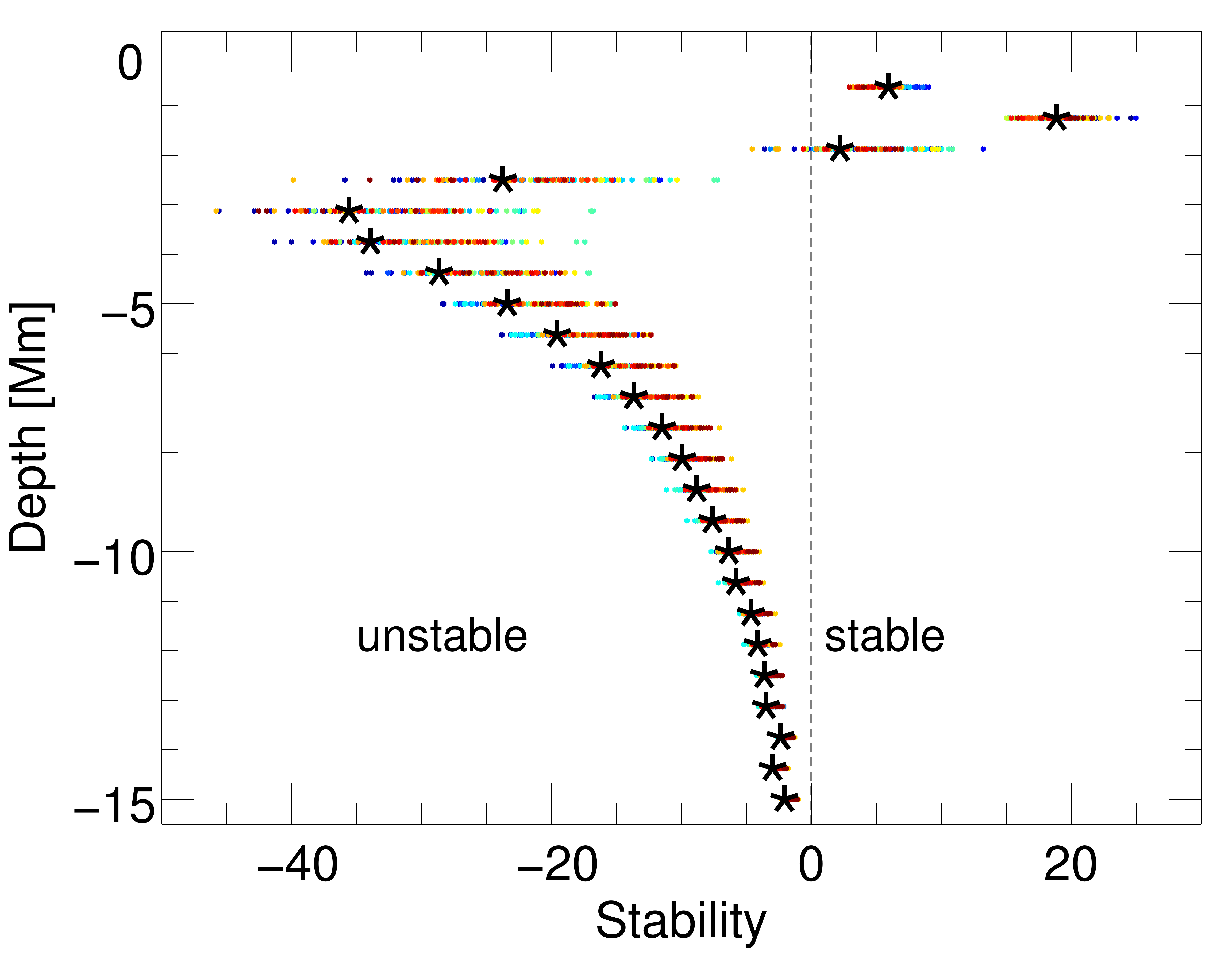}
\caption{Stability criterion of the sunspot flux tube at different depths time-averaged (black stars) over all 120 time steps (rainbow-coloured dots, blue to red in time). Positive values refer to a stable magnetic flux tube, and negative values represent an unstable configuration.}
\label{fig:stability}
\end{figure}

The stability of the sunspot at the different depth positions and for all time steps was studied by evaluating Eq.\,\eqref{eq:meyer_stability} with $g$\,=\,274\,m\,s\textsuperscript{-1}. When the difference on the left side of Eq.\,\eqref{eq:meyer_stability} is positive, then the flux tube is stable. The difference obtained for the different depths and time steps is visualised in Fig.\,\ref{fig:stability} as dots.\par
Down to $z$\,=\,-1.25\,Mm the difference is positive (see Fig.\,\ref{fig:stability}), and this means that the flux tube is stable. At $z$\,=\,-1.875\,Mm the time-averaged value is still positive, although individual time steps are negative, indicating an instability of the flux tube at these time steps. At greater depth, for $z$\,<\,-2.5\,Mm, the flux tube is found to be unstable for all time steps. The change from stable to unstable is almost always located at the same depth as the smallest radius of curvature, $R_\text{c}$, of the sunspot flux tube. At this depth the surface function has an inclination angle, $\chi$, between 42$^{\circ}$ and 62$^{\circ}$ with a temporal average of 46$^{\circ}$. The difference reaches a minimum at $z$\,=\,-3.125\,Mm. There, the flux tube is most unstable. With increasing depth, the negative difference approaches zero. However, the flux tube does not reach a stable configuration down to the lowest measurement point.\par
%
Fig.\,\ref{fig:stability} shows that the coloured dots for each depth do not follow the rainbow pattern. This means that no time-dependent relation of the flux-tube stability can be found.
\section{Discussion}\label{sec:disc}
We applied the stability criterion that was derived with a simple magneto-static model by \citet{Meyer_1977} in a current numerical simulation. We find that our simulated sunspot is stable in the upper 2\,Mm below the photosphere. Our analysis of the compactness of the flux tube shows that the compactness of the sunspot is constant at the surface. For layers deeper than $z$=-1.25\,Mm, the compactness decreases in time. \citet{Meyer_1977} have suspected that a sunspot flux tube could be stable up to a depth of 2\,Mm, although their magneto-static model only extended down to 0.650\,Mm below the photosphere. In these layers, the inclined magnetic field is predominately kept together by buoyancy forces. The density difference between the magnetic flux tube and the surroundings stabilises the sunspot.\par
In deeper layers, the stability criterion yields an unstable configuration, as shown in Fig.\,\ref{fig:stability}. This represents the increasing effect of fluting. The flux tube is least stable at around $z$\,=\,-3\,Mm, whereas the fastest decrease in compactness is found in deeper layers (see Fig.\,\ref{fig:decay_time}). We conjecture that the interchange process is damped within this region. Above $z$\,=\,-1.5\,Mm, the field lines are held together by buoyancy to a stable configuration. Magnetic tension affects the behaviour of the field lines in deeper layers. These non-local effects are not taken into account in the linear stability analysis. The field lines can move more freely at larger depths. Therefore the decrease in compactness is faster in layers below $z$\,=\,-3\,Mm. It proceeds most rapidly between $z$\,=\,-5\,Mm and $z$\,=\,-10\,Mm (see Fig.\,\ref{fig:decay_time}).\par
In even deeper layers, the decrease in compactness is found to proceed more slowly, and the instability of the flux tube approaches a marginally stable state (see Fig.\,\ref{fig:stability}). This behaviour is caused by a combination of two effects: (1) The convective timescale increases with depth, as described in Sec.\,\ref{sec:convtime} and shown in Fig.\,\ref{fig:convtime}. This leads to the slower action of destabilising forces. (2) The flux tube is more vertical and less strongly curved, which makes it less susceptible to fluting, an effect that has been predicted by \citet{Meyer_1977}.\par
The existence of the fluting stability alone is not sufficient for sunspot decay. In addition, turbulent flows are required that remove flux from the spot. \citet{Rempel_2015} found a stabilising effect of the moat cell that developed around the simulated sunspot. Below the penumbra, the spot is surrounded by an annulus of upflowing plasma that has a strongly reduced convective RMS velocity and is almost devoid of downflows. This strongly inhibits turbulent transport and in particular prevents the submergence of field lines below the penumbra, which would be required for a decay of the penumbra.\par
A non-monolithic sunspot flux tube was proposed by \citet{Parker_1979}. He described the individual field lines of a sunspot that are to be kept together in the photosphere. The observable monolithic structure would be divided into several smaller flux bundles approximately 1\,Mm below the visible photosphere.\par
\begin{figure}[!ht]
\centering 
\includegraphics[width=0.49\textwidth,height=\textheight,keepaspectratio]{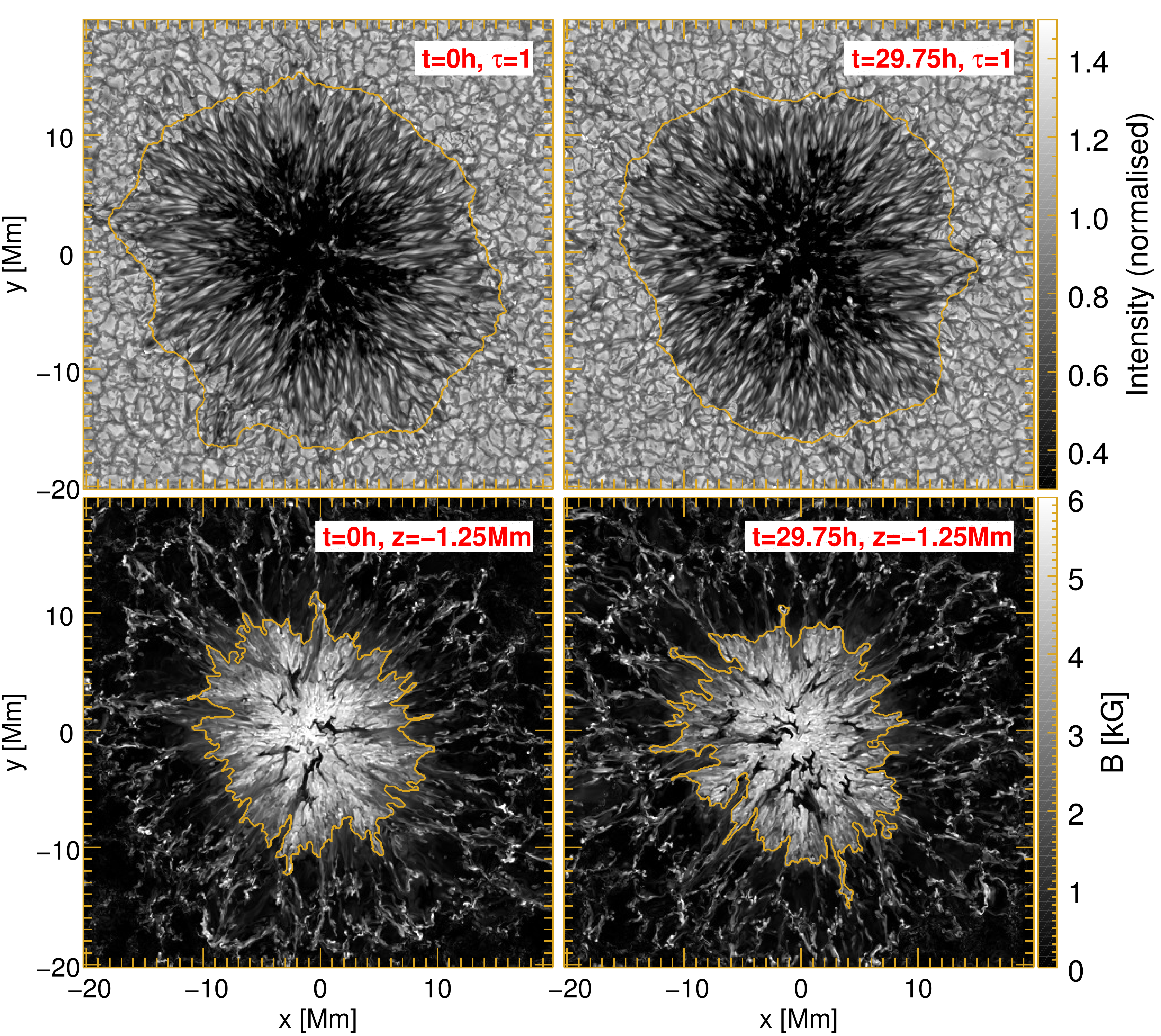}
\caption{Comparison of the sunspot at the first ($t$\,=\,0h, left) and last ($t$\,=\,29.75h, right) time step in intensity maps at $\tau$\,=1 (top row) and the magnetic field strength at a depth of 1.25\,Mm (bottom row). The outlined contours are determined with $I_{\mathrm{c}}=0.9I_{\mathrm{n}}$ and $B_{\mathrm{c}}=1669$\,G.}
\label{fig:evol_intB}
\end{figure}

The formation of intrusions at the sunspot boundary, which penetrate the flux tube below the visible surface, was also found by \citet{Panja_2021}. They calculated MHD simulations with different initial values for the vertical magnetic field at the bottom boundary of their simulation domain. After about $10$~hours, the sunspots simulated by \citet{Panja_2021} showed a fragmentation of the subsurface field that was substantially stronger than what we found here. This might be due to their more shallow domain (about 5.5 Mm), which cuts out the deeper part of the spot that is less susceptible to fluting. Despite this difference, they also find that the intrusions evolve fastest in depths below $z$\,=\,-5\,Mm. Another critical difference is that their setup did not impose a strongly inclined magnetic field at the top boundary. They found that the fluting instability promotes penumbra formation, which is not the case in our simulation. Switching to a potential field boundary condition leads to a naked spot without penumbra in about half an hour, as shown in \citet{Rempel_2015}. The fluting instability does lead to a growth of the penumbra over time at the expense of the umbra. This evolution was found in \citet{Rempel_2015} and can be seen in intensity maps in the top panels of Fig.\,\ref{fig:evol_intB}. This points in the same direction as the findings of \citet{Panja_2021}: The intrusions cause a conversion of magnetic flux from the umbra to the penumbra.\par
The observational study of the decay of a sunspot by \citet{Benko_2018} showed that the decrease in magnetic field strength within the umbra is one of the first observable effects of sunspot decay. In time, regions of weaker field are occupied by the penumbra or granulation. \citet{Benko_2018} concluded that the umbral area decreases during the decay phase of the sunspot. In our simulation the intrusions of field free plasma that are most prominent at a depth of a few Mm do lead to a visible imprint at the surface (see Fig.\,\ref{fig:evol_intB}). While we do not see fully developed light bridges, we find penumbral intrusions and clusters of higher umbral dot density associated with the subsurface fluting. This suggests that subsurface decay does not remain hidden from the surface.\par
We conjecture that the degradation of the sunspot at its outer surface will in time lead to the separation of the large magnetic flux tube into smaller ones. Light bridges that are observed as the onset of the decay could be caused by such intrusions. The weak magnetic field found within the sunspot in regions above $z$\,=\,-5\,Mm, at $t$\,>\,20\,h is the first indication of such a scenario. This agrees with the observations reported by \citet{McIntosh_1981} that the process that leads to sunspot decay should already act before the sunspot has fully developed. An interesting property of the simulated sunspot is that the total flux remains constant in spite of the subsurface decay, which can be attributed to the stabilising buoyancy of the penumbra and the stabilising effect of the moat cell. We conjecture that this confinement will eventually fail and at that time lead to a rapid disappearance of the already heavily corrugated spot. 
\section{Conclusion}\label{sec:conc}
We report the following main results from our study:
(1) The convective timescale (see Fig.\,\ref{fig:convtime}), increases with depth and reaches a value of 8\,hours at $z=-15$\,Mm, (2) the compactness of the sunspot decreases fastest at a depth around $z=-8$\,Mm (see Fig.\,\ref{fig:decay_time}), and (3) our stability analysis reveals that the sunspot is stable down to approximately $-2$\,Mm and most unstable around $z=-3$\,Mm (see Fig.\,\ref{fig:stability}). This leads to a consistent picture of sunspot stability and decay.\par
Our analysis shows that sunspot (in)stability and decay is an interplay between stabilising and destabilising forces as well as timescales. Buoyant inclined field lines just beneath the photosphere stabilise the sunspot and keep the field lines together. Non-local magnetic tension effects damp the fluting instability in the adjacent layers. In even deeper layers, the fluting instability steadily decreases the compactness of the flux tube. Thus, the spot eventually loses its coherency in layers between $z$\,=\,-5\,Mm and $z$\,=\,-10\,Mm. Deeper layers are less strongly affected because the destabilising force is weaker, convective timescales (time of change) increase substantially with depth, and the moat cell is stabilising.\par
We recall that this scenario of sunspot decay is based on simulated data. The described process occurs within the upper convection zone, a region hidden from real observations. For a more detailed clarification of the decay process, we might therefore have to compare characteristics of observations with those seen in the photospheric region of the simulated sunspot. Tracers of the degradation process that are visible in the photosphere could then be studied in greater detail in real observations. We propose to take light bridges as tracers for the indentations at the boundary of the flux tube below the solar surface into account. \citet{Panja_2021} reported that upflows within the intrusions become visible as light bridges at the surface. The role of MMFs in the sunspot decay process has been discussed in many studies (see Sec.\,\ref{sec:intro}). Their connection to the sunspot should be studied in simulations. In simulations we can take advantage of tracing the connection of their field lines below the surface. This might give insight into a possible role of these features in the interchange and degradation process of a sunspot.
\begin{acknowledgements}
We gratefully acknowledge fruitful discussions with Nazaret Bello Gonz\'alez and Markus Schmassmann, and we thank Rebecca Centeno Elliott for carefully reading and commenting on the manuscript. We appreciate the useful comments and suggestions of the anonymous referee. HS has been funded by the Deutsche Forschungsgemeinschaft, under grant No. RE 3282 and acknowledges financial support from the HAO visitor program. This material is based upon work supported by the National Center for Atmospheric Research, which is a major facility sponsored by the National Science Foundation under Cooperative Agreement No. 1852977.
\end{acknowledgements}
\bibliographystyle{aa}
\bibliography{bibfile_40199}
\begin{appendix}
\section{Determination of the flux tube boundary}\label{app:bcrit}
\begin{table*}[!h]
\begin{center}
\caption{Threshold levels $B_c$ and $B_{\text{eye}}$ for determining the boundary of the sunspot in maps of the magnetic field strength for defined depths $z$.}
\resizebox{.95\linewidth}{!}
{\begin{tabular}{l              c               c               c               c               c               c               c               c               c       }\hline
{\bf Depth [Mm]}        &       {\bf 0  }&      {\bf -0.625}    &       {\bf -1.25}  &       {\bf -1.875}    &       {\bf -2.5}      &       {\bf -3.125     }&      {\bf -3.75}  &       {\bf -4.375     }&      {\bf -5}         \\\hline
{\bf $B_{c}$ [G]}       &       600     &       977     &       1669    &       2203    &       2711    &       3139    &       3500    &       3806    &       4130    \\\hline
{\bf $B_{\text{eye}}$ [G]}      &       1000    &               &       1700    &               &       2000    &               &       2500    &               &       3500    \\\hline\\\hline
{\bf Depth [Mm]}        &       {\bf -5.625}    &       {\bf -6.25}     &       {\bf -6.875} &       {\bf -7.5}      &       {\bf -8.125     }&      {\bf -8.75}     &       {\bf -9.375} &       {\bf -10}       &       {\bf -10.625}   \\\hline
{\bf $B_{c}$ [G]}       &       4358    &       4547    &       4731    &       4782    &       4832    &       4928    &       5021    &       5036    &       5055    \\\hline
{\bf $B_{\text{eye}}$ [G]}      &               &       4000    &               &       4250    &               &       4500    &               &       4500    &               \\\hline\\\hline
{\bf Depth [Mm]}        &       {\bf -11.25}    &       {\bf -11.875}   &       {\bf -12.5}  &       {\bf -13.125}   &       {\bf -13.75}    &       {\bf -14.375}   &       {\bf -15}    &       {\bf -16.25}    &       {\bf -17.5}     \\\hline
{\bf $B_{c}$ [G]}       &       5104    &       5156    &       5084    &       5014    &       5017    &       5086    &       5100    &       5309    &       5509    \\\hline
{\bf $B_{\text{eye}}$ [G]}      &       4500    &               &       4500    &               &       4500    &               &       4500    &               &               \\\hline        
\end{tabular}\label{tab:bcrit}}
\end{center}
\end{table*}
\section{Compactness of the flux tube}\label{app:compactness}
\begin{figure*}[!hp]
\centering 
\includegraphics[height=0.96\textheight,width=\textwidth,keepaspectratio]{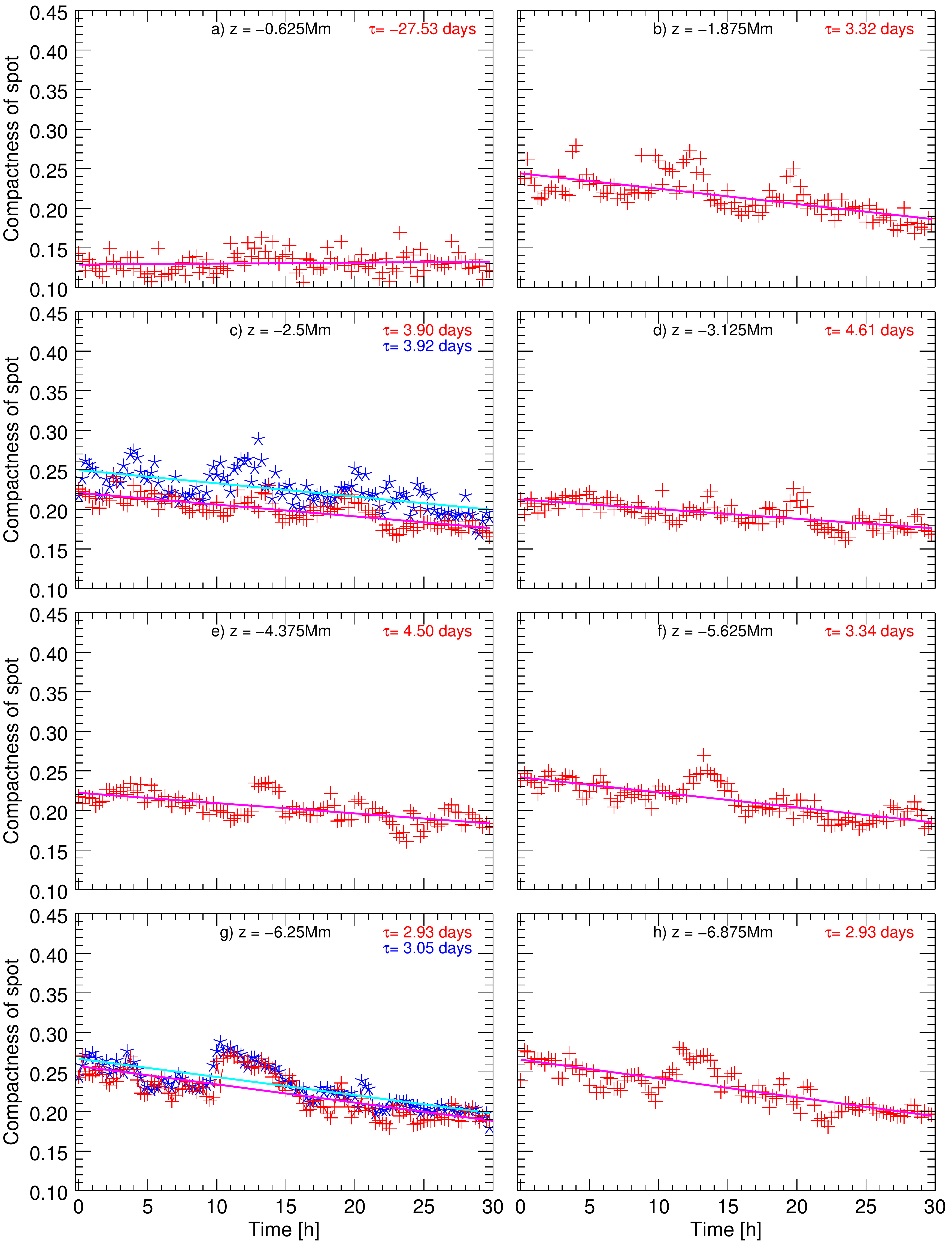}%
\caption{Compactness of the sunspot (normalised area-perimeter ratio) in time, shown at different depths. The boundary of the sunspot is determined with depth-dependent contour values $B_{c}(z)$ (red crosses) and $B_{\text{eye}(z)}$ (blue stars). Same as Fig.\,\ref{fig:compactness}.}
\label{fig:compactness_app1}
\end{figure*}
\begin{figure*}[!hp]
\centering 
\includegraphics[height=0.96\textheight,width=\textwidth,keepaspectratio]{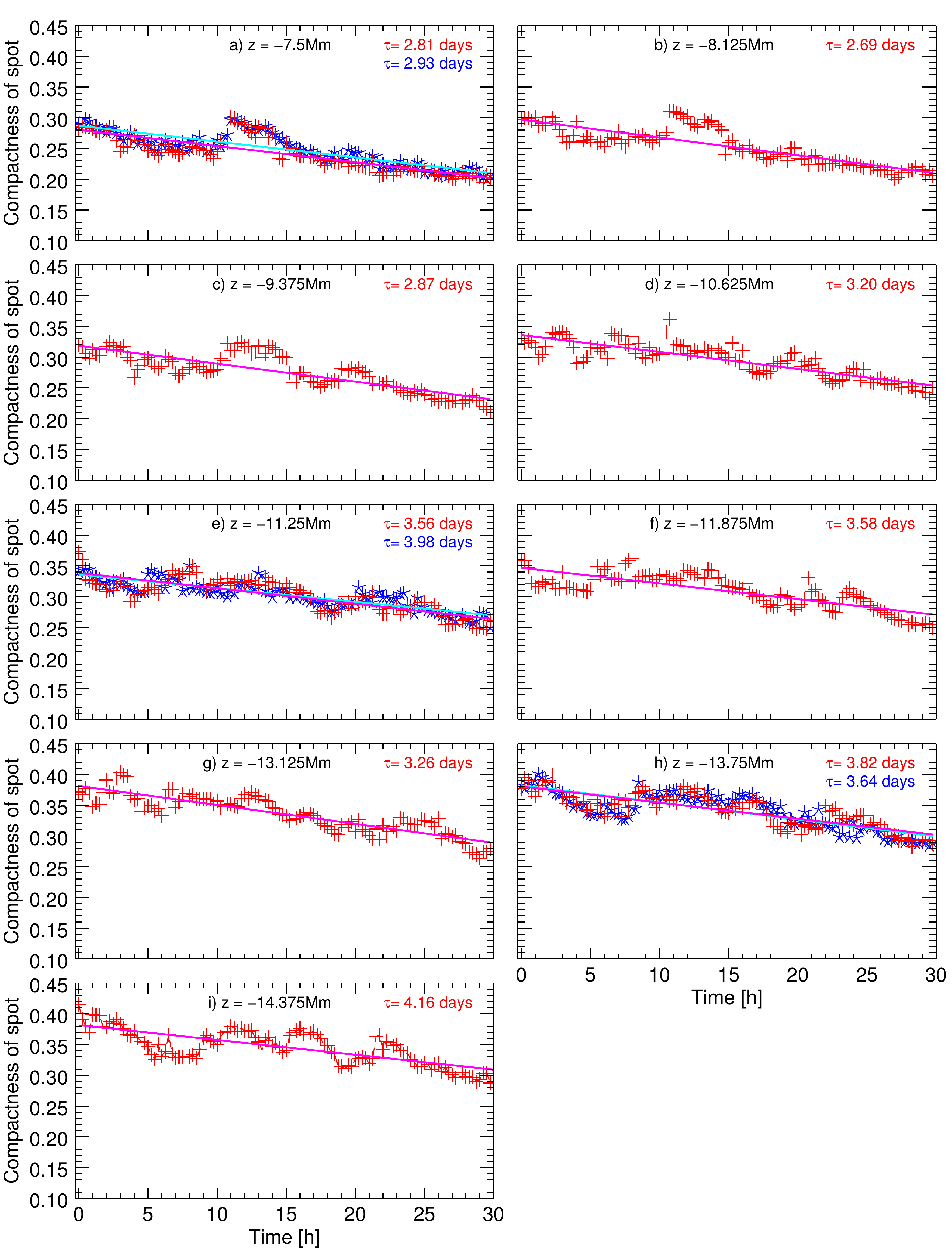}%
\caption{Same as Fig.\,\ref{fig:compactness_app1}, but for different depths.}
\label{fig:compactness_app2}
\end{figure*}
\end{appendix}
\end{document}